\documentclass[preprint,superscriptaddress,notitlepage,endfloat]{revtex4-2}
\usepackage{graphicx,bm,mathtools,color}
\usepackage[pdfstartview=FitH, CJKbookmarks=true, bookmarksnumbered=true, bookmarksopen=true, colorlinks, pdfborder=001, linkcolor=blue, anchorcolor=blue, citecolor=blue]{hyperref}
\usepackage{ulem}
\usepackage{verbatim}
\usepackage{upgreek}
\usepackage[justification=Justified]{caption}

\newcommand{\del}[1]{\iffalse #1 \fi}
\newcommand{\be}{\begin{equation}}
	\newcommand{\ee}{\end{equation}}

\newcommand{\LNO}{La$_3$Ni$_2$O$_{7-\updelta}$}
\newcommand{\LN}{La$_3$Ni$_2$O$_7$}

\newcommand{\Tc}{$T_{\rm c}$ }
	
\begin{document}

	\title{High-temperature superconductivity with zero-resistance and strange metal behaviour in \LNO}

	\author{Yanan Zhang}
	\thanks{These authors contributed equally to this work.}
	\author{Dajun Su}
	\thanks{These authors contributed equally to this work.}
	\affiliation  {Center for Correlated Matter and School of Physics, Zhejiang University, Hangzhou 310058, China}

	\author{Yanen Huang}
	\affiliation  {Center for Correlated Matter and School of Physics, Zhejiang University, Hangzhou 310058, China}

	\author{Zhaoyang Shan}
	\affiliation  {Center for Correlated Matter and School of Physics, Zhejiang University, Hangzhou 310058, China}

	\author{Hualei Sun}
	\affiliation  {School of Science, Sun Yat-sen University, Shenzhen 518107, China}

	\author{Mengwu Huo}
	\affiliation  {Center for Neutron Science and Technology, Guangdong Provincial Key Laboratory of Magnetoelectric Physics and Devices, School of Physics, Sun Yat-Sen University, Guangzhou, Guangdong 510275, China}


	\author{Kaixin Ye}
	\affiliation  {Center for Correlated Matter and School of Physics, Zhejiang University, Hangzhou 310058, China}

	\author{Jiawen Zhang}
	\affiliation  {Center for Correlated Matter and School of Physics, Zhejiang University, Hangzhou 310058, China}

	\author{Zihan Yang}
	\affiliation  {Center for Correlated Matter and School of Physics, Zhejiang University, Hangzhou 310058, China}

	\author{Yongkang Xu}
	\affiliation  {Center for Correlated Matter and School of Physics, Zhejiang University, Hangzhou 310058, China}

	\author{Yi Su}
	\affiliation  {Center for Correlated Matter and School of Physics, Zhejiang University, Hangzhou 310058, China}

	\author{Rui Li}
	\affiliation  {Center for Correlated Matter and School of Physics, Zhejiang University, Hangzhou 310058, China}



	\author{Michael Smidman}
	\affiliation  {Center for Correlated Matter and School of Physics, Zhejiang University, Hangzhou 310058, China}

	\author{Meng Wang}
	\email[Corresponding author: ]{wangmeng5@mail.sysu.edu.cn}
	\affiliation  {Center for Neutron Science and Technology, Guangdong Provincial Key Laboratory of Magnetoelectric Physics and Devices, School of Physics, Sun Yat-Sen University, Guangzhou, Guangdong 510275, China}

	\author{Lin Jiao}
	\email[Corresponding author: ]{lin.jiao@zju.edu.cn}
	\affiliation  {Center for Correlated Matter and School of Physics, Zhejiang University, Hangzhou 310058, China}

	\author{Huiqiu Yuan}
	\email[Corresponding author: ]{hqyuan@zju.edu.cn}
	\affiliation{Center for Correlated Matter and School of Physics, Zhejiang University, Hangzhou 310058, China}
	\affiliation  {State Key Laboratory of Silicon and Advanced Semiconductor Materials, Zhejiang University, Hangzhou 310058, China}
	\affiliation  {Zhejiang Province Key Laboratory of Quantum Technology and Device,
		School of Physics, Zhejiang University, Hangzhou 310058, China}
	\affiliation  {Collaborative Innovation Center of Advanced Microstructures, Nanjing 210093, China}

	\date{\today}

	\begin{abstract}
\textbf{Recently signatures of superconductivity were observed close to 80 K in \LN\ under pressure \cite{Wang2023arxiv}. This discovery positions \LN\ as the first bulk nickelate with high-temperature superconductivity, but the lack of zero resistance presents a significant drawback for validating the findings. Here we report pressure measurements up to over 30 GPa using a liquid pressure medium and show that single crystals of \LNO\ do exhibit zero resistance.
We find that \LNO\  remains metallic under applied pressures, suggesting the absence of a metal-insulator transition proximate to the superconductivity. Analysis of the normal state $T$-linear resistance suggests an intricate link between this strange metal behaviour and superconductivity, whereby at high pressures both the linear resistance coefficient and superconducting transition are slowly suppressed by pressure, while at intermediate pressures both the superconductivity and strange metal behaviour appear disrupted, possibly due to a nearby structural instability. The association between strange metal behaviour and high-temperature superconductivity is very much in line with diverse classes of unconventional superconductors, including the cuprates and Fe-based superconductors \cite{Cooper2009,Jiang2023NP,Taillefer2010,Greene2020,Phillips2022}. Understanding the superconductivity of \LNO\ evidently requires further revealing the interplay of strange metal behaviour, superconductivity, as well as possible competing electronic or structural phases.}
	\end{abstract}

	\maketitle

Quasi-two-dimensional layered transition metal oxide structures are a common motif for high-temperature unconventional superconductivity~\cite{Norman2014,stewart2017}, yet this elusive phenomenon has only been realized in a handful of material classes, primarily the Cu-based cuprate \cite{Muller1986, Chu1987} and Fe-based  superconductors \cite{Hosono2008,Chen2008}. Very recently, signatures of superconductivity up to nearly 80~K were revealed in bulk samples of the nickelate \LN\ above 14~GPa \cite{Wang2023arxiv}. This is in contrast to the infinite-layer nickelates whereby superconductivity with \Tc of 5-15~K is found in thin-films \cite{,Li2019nature,Ariando2020PRL,Harold2020PRM,Ariando2022scienceadvance,Ding2023nature}, increasing to 31~K under pressure \cite{JGCheng2022NC}, but superconductivity is not observed in bulk samples \cite{WenHaiHu2020CM,Phelan2020PRM}. While the Ni$^+$ of the infinite-layer nickelates has the same $3d^9$ configuration as the Cu$^{2+}$ of the cuprates, the presence of apical oxygens in \LN\ leads to a formal $3d^{7.5}$ configuration. It is suggested that interlayer coupling between Ni $3d_{z^2}$  and apical oxygen $p$ orbitals at high pressures leads to partially occupied $3d_{x^2-y^2}$ orbitals \cite{Wang2023arxiv}, \del{resembling the physics of the cuprates \cite{Lee2006,Keimer2015},} but the relevant orbitals and pairing instabilities for the superconductivity remain to be clarified \cite{Nakata,YaoDaoXin2023,Elbio2023ARXIV,HuJiangping2023ARXIV,Eremin2023ARXIV,WANGQIANGHUA2023ARXIV,Kazuhiko2023ARXIV,ZhangGuangMing2023ARXIV,LeonovPRB2023,WenHaiHuARXIV2023,Philipp2023ARXIV,YangYifengARXIV2023}.

\begin{figure}[h]
\includegraphics[width=0.7\columnwidth]{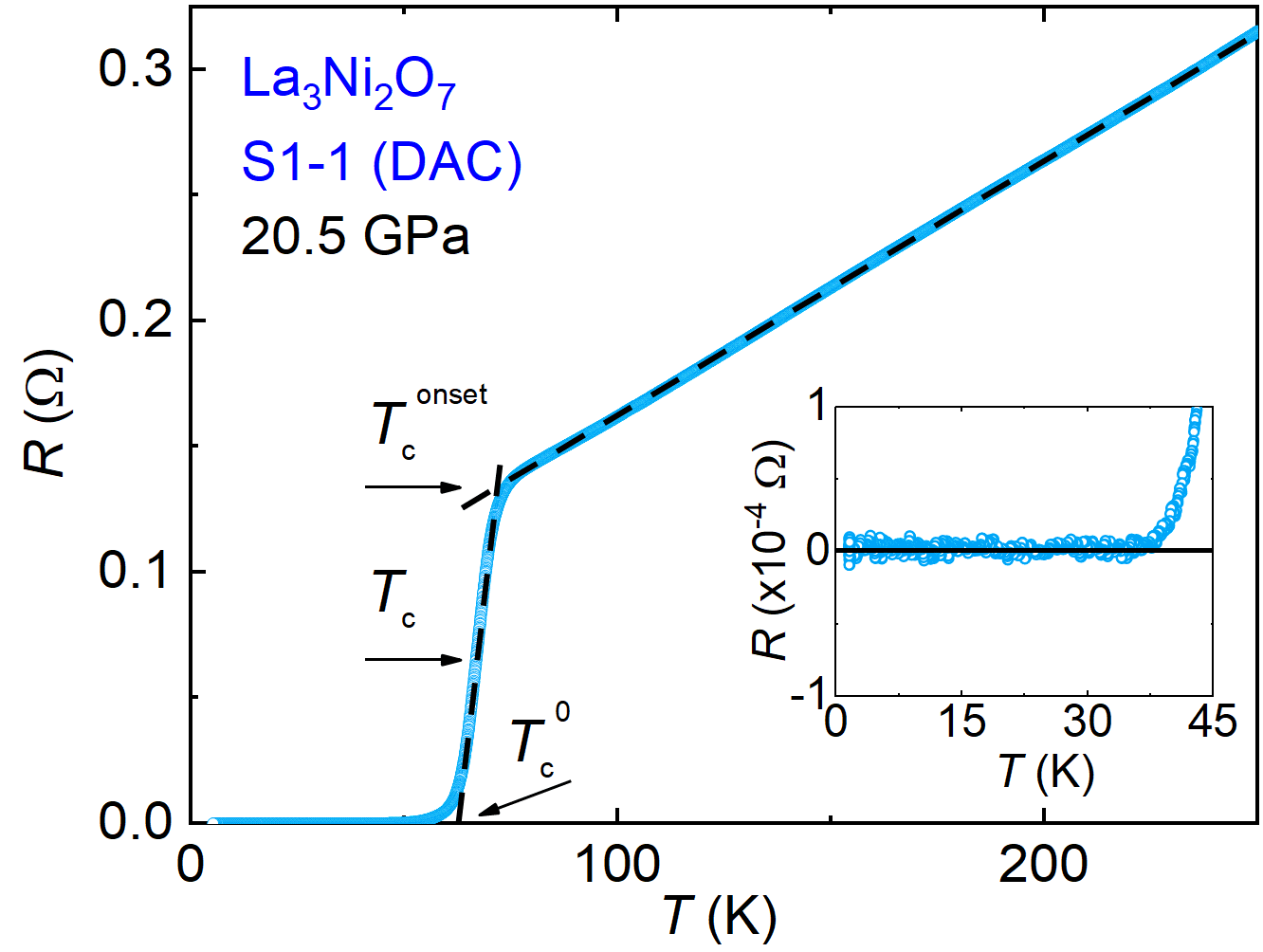}
\caption{\textbf{Zero-resistance in \LNO\ at 20.5~GPa.} Temperature dependence of the resistance $R(T)$ of \LNO\ under a pressure of 20.5~GPa, showing a clear superconducting transition onsetting below 66~K, and reaching zero resistance just below 40~K. The normal state  $R(T)$ above \Tc shows a linear temperature dependence corresponding to strange metal behaviour. The construction used to determine $T_c^{onset}$, $T_c$, and $T_c^0$ is illustrated in the figure. The inset shows an enlargement of the resistance below \Tc, clearly demonstrating that zero-resistance is reached.}
\label{Fig1}
\end{figure}

	Although signatures of high-temperature superconductivity are reported from a drop of the electrical resistivity, as well as a diamagnetic response of the ac susceptibility \cite{Wang2023arxiv}, crucially zero resistance was not reached in electrical resistivity measurements, which is a vital criterion for establishing the occurrence of superconductivity. Here we report electrical resistance measurements of \LNO\ under pressure, which were performed in both a piston-cylinder cell (PCC) and a diamond anvil cell (DAC) (inset of Extended Data Fig. 2), where the use of a liquid pressure-transmitting-medium leads to a greatly improved hydrostaticity. As displayed in Fig.~\ref{Fig1}, at a pressure of 20.5~GPa a clear superconducting transition is observed in $R(T)$ which onsets below 66~K, and  close to 40~K zero resistance is reached within the noise level of the measurement system. Above $T_{\rm c}$, $R(T)$ exhibits a linear temperature dependence, characteristic of strange metal behaviour \cite{Wang2023arxiv}. Such a zero-resistance state at temperatures around 40~K clearly demonstrates high-temperature superconductivity in La$_3$Ni$_2$O$_7$. Robust signatures of superconductivity are also manifested in measurements of the resistivity of another sample when measured with different currents, as well as $I-V$ curve measurements (Extended Data Fig. 3). Linear $I-V$ curves are observed above \Tc as expected in the normal state, while the curves are flat at low temperatures and currents, yielding a superconducting critical current of $\sim$850 A/cm$^2$ ($T$ = 1.5 K and $P$ = 16.6 GPa).

	In order to confirm the reproducibility of the zero resistance, several additional samples were measured, among which zero-resistance were observed in three samples, while in three others there is either a drop in resistance that does not reach zero or weakly insulating behaviour (Extended Data Fig. 4). A comparison of energy dispersive x-ray spectroscopy results shows that the \LNO\ samples exhibiting zero resistance have a composition closer to the stoichiometric ratio of La:Ni:O=3:2:7, with a relatively small variation of the composition across the sample (Extended Data Figs. 4 and 5, Table S1). On the other hand, samples without zero resistance are further from stoichiometry with a larger compositional inhomogeneity. Nevertheless, the precise compositional range of the superconductivity, as well as the optimal composition, remain to be determined.
		Note that all the samples in this study are from the same growth as those in Ref.~\cite{Wang2023arxiv}, where zero-resistance was not observed, but those measurements were performed either with no pressure-transmitting medium, or with a solid medium, suggesting that more hydrostatic pressures realized with a liquid medium are crucial for its observation. Furthermore, the small sample dimensions (of order 10-100~$\mu$m) used in our study means that  homogeneous samples with only  small variations in the composition across the sample can be measured, allowing for the observation of a sharp superconducting transition with zero resistance.

\begin{figure}[h]
\includegraphics[width=0.55\columnwidth]{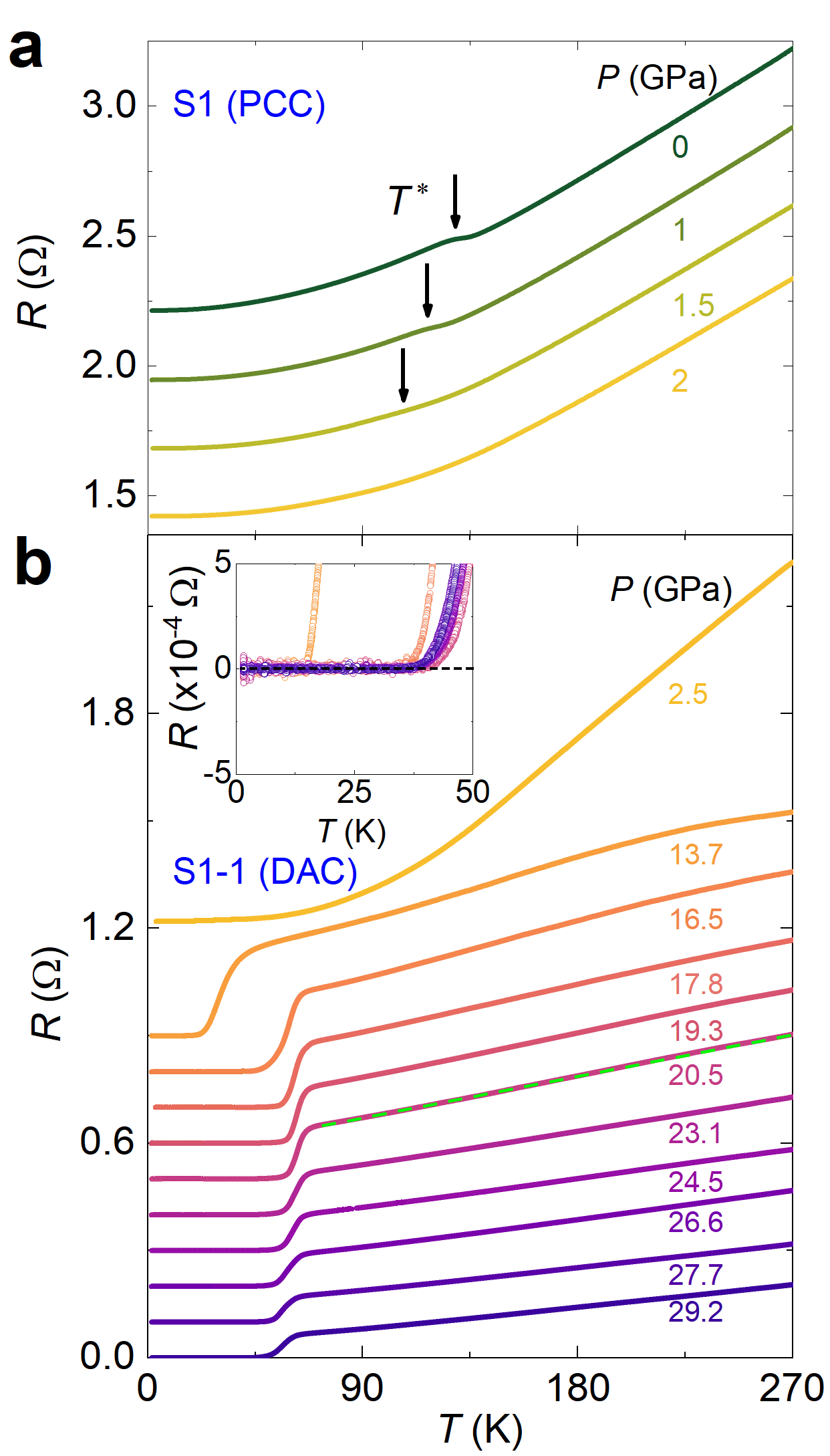}
\caption{\textbf{Temperature dependent resistance of \LNO\ under pressure.} \textbf{a,}  $R(T)$ of \LNO measured at both ambient pressure and under various hydrostatic pressures up to 2~GPa in a PCC. The kink corresponding to the possible DW transition \cite{Wang2022sciencechina} is denoted as $T^*$. \textbf{b,} $R(T)$ measured under higher pressures from 2.5~GPa to 29.2~GPa in a DAC. Dashed green line show the linear temperature dependence in the normal state. Note that while the curves are equally shifted vertically for clarity, the data at 13.7~GPa and above all reach zero-resistance below the superconducting transition, as demonstrated in the inset.}
			\label{Fig2}
		\end{figure}

		The temperature dependence of the resistance $R(T)$ is displayed in Fig.~\ref{Fig2} for various pressures up to 29.2~GPa. At ambient pressure, $R(T)$ is metallic across the entire temperature range, and there is an anomaly around $T^*=130$~K. This temperature scale is close to the possible density wave (DW) transition \del{observed} suggested in previous studies \cite{Wang2022sciencechina, Hundley2001PRB}. When a small pressure is applied, $R(T)$ remains metallic while $T^*$ decreases slightly at 1.0~GPa, and is not detected at 2.0~GPa.
		\LNO\ being consistently metallic under pressure is in contrast to Ref.~\cite{Wang2023arxiv} where weakly insulating behaviour is induced by a small pressure, which in turn disappears following the structural transition above 10~GPa.
		Our findings therefore are much more in line with reports showing that insulating behaviour in \LNO\ is associated with oxygen deficient rather than stoichiometric samples \cite{ZHANG1994,Hundley2001PRB,Masatoshi1995JPSJ}. Note that the DAC  measurements of the low-pressure structural phase ($P$ $<$ 13~GPa) exhibit a large contact resistance between the sample and Au leads, which increases significantly with decreasing temperature, and could give rise to spurious insulating behaviour \cite{Blankenship1988YBCO} (Extended Data Fig. 6). This sizeable contact resistance is absent in DAC measurements performed above 13 GPa, and it is avoided in our piston-cylinder cell measurements below 2.5 GPa by increasing the area of the contacts and by avoiding heating the sample.

		At pressures above 13~GPa, superconducting transitions are observed in $R(T)$, which reach zero resistance below the transition (inset of Fig.~\ref{Fig2}b). Upon increasing the pressure, the onset of the superconducting transition $T^{\rm onset}_{\rm c}$ increases from 37.5~K at 13.7~GPa to a maximum of 66~K at 20.5~GPa. Further increasing the pressure (up to 30~GPa) results in a gradual decrease of $T^{\rm onset}_{\rm c}$, with d$T_{\rm c}$/d$P$ $\approx$ $-$0.8~K/GPa. A similar pressure dependence of $T^{\rm onset}_{\rm c}$ is measured in other samples (Extended Data Fig. 4), where the broader transitions \del{is} in some samples are likely a consequence of sample inhomogeneity.  Strange metal behaviour is also observed  above around 13.7 GPa, which corresponds to where there is a linear temperature dependence of $R(T)$. At 13.7~GPa, where there is evidence of a possible structural transition at higher temperatures (Extended Data Fig. 2), the linear $R(T)$ extends only up to around 100~K, but at 20.5~GPa this reaches at least 270~K. From analyzing with $R$($T$) = $R$(0) + $A^{\prime}T$ (dashed green line in Fig.~\ref{Fig2}), it is found that the $T$-linear coefficient $A^{\prime}$ decreases with increasing pressure, as discussed below.

\begin{figure}[h]
				\includegraphics[width=0.8\columnwidth]{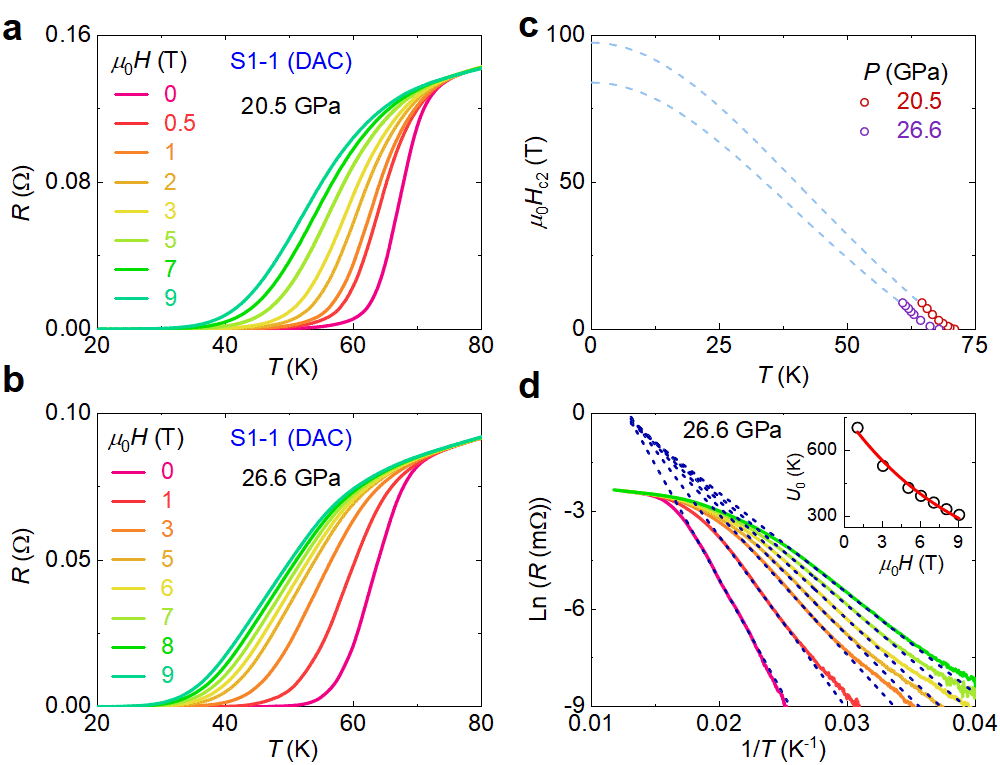}
			\caption{\textbf{Resistance of \LNO\ in magnetic fields and thermally activated flux flow.} $R(T)$ of \LNO\ is displayed for \textbf{a,} 20.5~GPa, and  \textbf{b,} 26.6~GPa, under various applied fields up to 9~T.  \textbf{c,} Temperature dependence of the upper critical fields at the two pressures, where dashed lines show the results of fitting with a Ginzburg-Landau model. \textbf{d,} Arrhenius plot of $R(T)$ in the region of the superconducting transition at 26.6~GPa in various magnetic fields. The dashed lines are linear fittings to Ln($R$) versus 1/$T$. The inset shows the thermal activation energy $U_0(H)$ obtained from the slopes of the plot. The red curve represents $U_0(H)$ $\propto$ $H^{-0.12}$. Panels \textbf{b} and \textbf{d} share the same color code.}
			\label{Fig3}
		\end{figure}

		Figures~\ref{Fig3}a and \ref{Fig3}b display $R(T)$ in different applied magnetic fields under pressures of 20.5 and 26.6~GPa, respectively. The derived upper critical fields $H_{c2}(T)$ (defined from $T_c^{onset}$) versus temperature are shown in Fig~\ref{Fig3}c (see also Extended Data Fig. 7), where the initial slopes d$\mu_0H_{c2}(T)$/d$T$ are 0.55~T/K and 0.57~T/K for 20.5~GPa and 26.6~GPa, respectively. By fitting $H_{c2}(T)$ with a Ginzburg-Landau model, respective zero-temperature values of 97~T and 83~T are obtained, from which  coherence lengths of 1.84 nm and 1.99 nm are deduced using $\mu_0H_{c2}$(0) = $\Phi_0$/2$\pi\xi_{GL}^2$. Moreover, there is a gradual broadening of the transition with increasing field, indicating thermally activated flux flow (TAFF) in \LNO. As shown by the Arrhenius plots in Fig~\ref{Fig3}d, the resistance data near the transition can be analyzed based on the simplified TAFF model: $U_0$($H$) = $-$dLn$R$/d(1/$T$), where $U_0$ is the thermal activation energy \cite{Vinokur1994RMP}. It is noted that the Arrhenius scaling holds over 3-4 orders until the signal reaches the noise floor of our instrument. These behaviours very much resemble iron-pnictide \cite{Wen2008} and chalcogenides \cite{Jiao2012}, as well as high $T_c$ cuprates \cite{Waszczak1990PRB}. Fitting to the TAFF model  derives an activation energy of $U_0$(1~T) = 702~K, which is much smaller than that for many cuprate and Fe-based superconductors, reflecting relatively weak pinning forces in \LNO. Upon increasing the applied magnetic field, $U_0$($H$) follows a power law behaviour ($\propto H^{-n}$), with a small exponent of $n$ = 0.12, indicating a weak field-dependence of $U_0$.

\begin{figure}[h]
				\includegraphics[width=0.45\columnwidth]{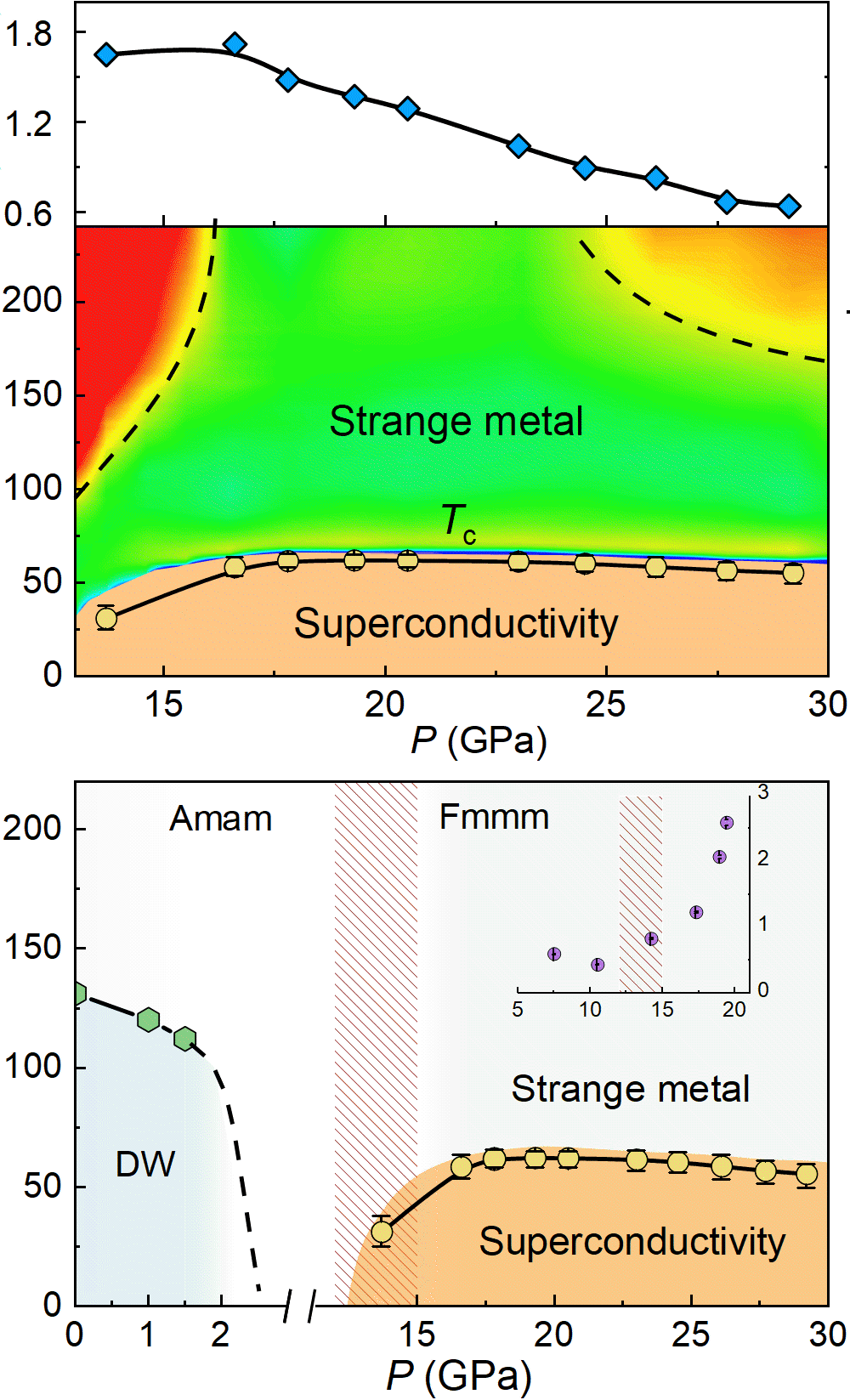}
			\caption{\textbf{Phase diagram of superconductivity and strange metal behaviour in \LNO.} \textbf{a,} Evolution of the $T$-linear coefficient $A^{\prime}$  versus pressure. \textbf{b,} Temperature-pressure phase diagram based on the resistance measurements in this work.  The color plot highlights the extent of the strange metal behaviour, where the colour represents the fractional deviation of $R(T)$ from fits to a linear temperature dependence $\left(R-R_n\right) / R_n(\%)$, where $R_n$ = $R_0$ + $A^{\prime}T$. The two black dashed lines highlight where the fractional deviation is around $\pm$2\%. \textbf{c,} Phase diagram summarizing the low and high pressure behaviours, where the former has a probable density wave (DW) transition, that is suppressed by 2~GPa of pressure. At around 10 GPa there is a change of crystal structure from one with space group $Amam$ to a structure with $Fmmm$ \cite{Wang2023arxiv}, as highlighted by the hashed area in the main panel and  inset, and at higher pressures there is the emergence of the high-temperature superconductivity and strange metal phases. The error bars for $T_c$ reflect the superconducting transition width, defined as $T_c^{onset}$ $-$ $T_c^0$. The inset shows the evolution of the reciprocal of the Hall coefficient 1/$R_H$ with pressure at 80 K.}
			\label{Fig4}
		\end{figure}

		The temperature-pressure phase diagram in Fig.~\ref{Fig4} shows the main results from the current pressure study. Superconductivity with zero-resistance appears above 13~GPa, and is rapidly enhanced by pressure, reaching a maximum \Tc of $\sim$62~K at 20.5~GPa, followed by a gradual decrease upon further pressure increases. The strange metal behaviour in the normal state is also highlighted by the color plot, in which the green regions correspond to where there is close to a $T$-linear resistance, while the pressure dependence of the $T$-linear coefficient $A^{\prime}$ is shown in the upper panel.  At 13.7 GPa, there is a relatively narrow temperature range over which there may be a $T$-linear $R(T)$ (up to around 100~K), and \Tc is correspondingly lower. On the other hand, the significant enhancement of \Tc at 16~GPa coincides with a marked expansion of the strange metal region, where the temperature range of the $T$-linear behaviour is largest near to the pressure where \Tc is a maximum, at which there is only a 1\% deviation from linearity. At higher pressures, both $A^{\prime}$ and \Tc decrease with pressure, and the temperature range of the strange metal region is again reduced, as shown by the dashed lines in Fig.~\ref{Fig4}b.

		Our observations suggest a close relationship between the strange metal behaviour and  high-temperature superconductivity in \LNO, where the decrease of \Tc with pressure is associated with a reduction of $A^{\prime}$. On the other hand, at 13.7~GPa there does not appear to be a correspondingly reduced $A^{\prime}$, but at this pressure a weak signature of a possible structural transition is detected at higher temperatures (Extended Data Fig. 2), which appears to disrupt both the superconductivity and strange metal state. This suggests that understanding the superconductivity  of \LNO\ requires revealing the various competing structural and electronic phases, which are highlighted when the phase diagram is extended to low pressures in Fig.~\ref{Fig4}c. Here a possible DW transition is suppressed by a moderate pressure, while at higher pressures there is a transition to a structural phase favouring both high-temperature superconductivity and strange metal behaviour.

		At this pressure-induced structural transition, there is a change from a low-pressure structure with tilted NiO$_6$ octahedra (space group $Amam$) \cite{Ling2000}, shown in Extended Data Fig. 1, to one with untilted octahedra (space group $Fmmm$) \cite{Wang2023arxiv}, and it is proposed that this corresponds to a change from the $3d_{z^2}$ orbitals being fully localized in the low pressure region, to a metallization of the  $3d_{z^2}$ bonding bands at high pressures \cite{Wang2023arxiv}. Our Hall resistivity ($\rho_{xy}$) measurements of \LNO\ under pressure (Extended Data Fig. 8) are consistent with this scenario, where in the context of there being both electron and hole Fermi surfaces, the linear $\rho_{xy}$ with a positive Hall coefficient $R_H$ suggests either predominantly hole carriers, or compensation between electron and hole Fermi surfaces. Above 15~GPa, we find a marked increase of $1/R_H$ at 80~K (which is above $T_c^{onset}$) (see the inset of Fig.~\ref{Fig4}c), as expected from an increased hole concentration arising as a result of the metallization of $3d_{z^2}$ bonding bands due to enhanced inter-layer $\sigma$-bond coupling through an inner apical oxygen. Interestingly, a high-temperature structural phase transition is observed  at ambient pressure in \LNO\ at $T_B\sim550$~K \cite{Sasaki1997}, below which it was also proposed that there is a localization of the $3d_{z^2}$ band. This suggests that the transition below $T_B$ at ambient pressure, and that at 13.7~GPa shown in Extended Data Fig. 2, may correspond to the same structural phase transition that terminates in the vicinity of the latter pressure. The existence of such a transition line in the $T-P$ phase diagram needs to be confirmed by further studies under pressure, and would indicate that the stablization of the $Fmmm$ structure is a highly promising route for realizing bulk superconductivity in the nickelates at ambient pressure.

		The rich interplay of the different ground states is very much in line with the cuprates and Fe-based superconductors \cite{Fernandes2014,Proust2019}. In particular, the initimate link between the strange metal phase and high-temperature superconductivity in \LNO\ is also reminiscent of the enigmatic relationship between these phenomena in the Cu-based, Fe-based, and infinite-layer nickelate superconductors \cite{Cooper2009,Jin2011,Taillefer2010,Licciardello2019,Zhao2022,Jiang2023NP,Taillefer2010,Greene2020,Lee2023}, whereby superconducting properties such as \Tc and the superfluid density are often correlated with $A^{\prime}$,  reaching a peak at optimal doping. Since Landau quasiparticles are anticipated to be absent in the strange metal phase, these in turn suggest that unconventional superconductivity could emerge via non-quasiparticle states \cite{Phillips2022}. As such, these results hint at an unconventional nature of the high-temperature superconductivity in \LNO, while underscoring the close association between strange metals and superconductivity across various materials systems, highlighting the nickelate superconductors as a new platform for examining this interplay.

		\textit{Note: In recent preprints \cite{JGChen2023, polyChengJGARXIV2023}, zero-resistance was observed in \LNO\ under pressure using a liquid-transmitting medium. Polycrystalline \LNO\ shares similar superconducting properties to the single crystalline \LNO\  reported in this work.}

		\clearpage

		\noindent \textbf{Acknowledgments}
	We acknowledge fruitful discussions with Chao Cao, Haiqing Lin, Frank Steglich, and Guangming Zhang. We also thank Shilong Zhang and Yingying Peng for sharing their unpublished single-crystal XRD results, as well as Zehao Dong, Yayu Wang, and Zhen Chen for sharing their unpublished TEM results.
	Work at Zhejiang University was supported by the National Key R\&D Program of China (Grant No. 2022YFA1402200 and No. 2023YFA1406100), the Key R\&D Program of Zhejiang Province, China (Grant No. 2021C01002), the National Natural Science Foundation of China (Grants No. 12274364, No. 12034017, No. 12174332, No. 12204408 and No. 12222410), and the Zhejiang Provincial Natural Science Foundation of China (Grant No. LR22A040002). L.J. was supported by “the Fundamental Research Funds for the Central Universities” (Grant No. 226-2024-00039). Work at Sun Yat-sen University was supported by the National Natural Science Foundation of China (Grant No. 12174454), Guangdong Basic and Applied Basic Research Funds (Grant No. 2021B1515120015), Guangzhou Basic and Applied Basic Research Funds (Grant No. 202201011123), and Guangdong Provincial Key Laboratory of Magnetoelectric Physics and Devices (Grant No. 2022B1212010008).
	\\
		\\
		\\
		\textbf{Additional information} Correspondence and requests for materials should be addressed to H. Q. Yuan (hqyuan@zju.edu.cn), L. Jiao (lin.jiao@zju.edu.cn), or M. Wang (wangmeng5@mail.sysu.edu.cn).
		\\
		\\
		\textbf{Author contributions}
		H.Y. and L.J. conceived the experiments. The single crystals were provided by H.S., M.H., and M.W.. Y.Z., D.S., Y.H., Z.S., K.Y., J.Z., Z.Y., Y.X., Y.S. and R.L. performed the transport, Laue, SEM and EDX measurements. 
  M.S., Y.Z., L.J. and H.Y. wrote the paper with input from all authors.
		\\
		\textbf{Competing financial interests} The authors declare no competing financial interests.

\clearpage

\clearpage
\clearpage
\clearpage

\clearpage
\noindent \textbf{Methods} \\
\noindent Polycrystalline rods of \LNO\ were prepared via a solid-state reaction, utilizing La$_2$O$_3$ (99.99$\%$ purity) and NiO (99.9$\%$ purity) in a 3:4 molar ratio. The thoroughly ground mixture was hydrostatically pressed into rods measuring 6~mm in diameter and 80~mm in length, followed by sintering at 1400$^{\circ}$C for 12 or 48 hours. Single crystals of \LNO\ were grown using a high pressure high temperature optical floating-zone furnace (HKZ, SciDre, Dresden) under an oxygen pressure of 15 bar and using a 5~kW Xenon arc. The travelling speed of the seed was 3 mm/h during the growth. The seed and feed rods were rotated in opposite directions at speeds of 15 and 10 rpm, respectively. The resulting growth consists of many  1-3 millimeter-sized single crystals. However, owing to robust interlayer coupling, cleaved single crystals may not consistently exhibit a uniform surface. In this work, we studied ten samples cut from the same growth. Samples S1, S2, and S3 are larger samples cut from the growth, and were measured in a piston-cylinder cell, while S1-1, S1-2, S1-3, S2-1, S2-2, S2-3 and S2-4, were smaller samples cut from  S1 and S2, and were measured in a diamond anvil cell. The data presented in the main text are from measurements of S1 and S1-1. The single crystalline nature and orientation of samples S1, S2, and S3 were confirmed using x-ray Laue measurements shown in Extended Data Fig. 7. The temperature dependence of the magnetic susceptibility of the three samples are also displayed in this figure for fields perpendicular to the $c$-axis, which all show similar behaviour, and also correspond well to the data in Ref.~\cite{LiuZ} for this field direction.

Measurements under applied pressures up to 2.0~GPa were carried out utilizing a piston-cylinder-type pressure cell  with Daphne 7373 used as the pressure-transmitting medium. The applied pressure was determined by the shift in $T_{\rm c}$ of a high-quality Pb single crystal~\cite{Eiling}.  For measurements at pressures above 2.0~GPa, the samples were polished and cut to approximate dimensions $120 \times 80 \times 20~\mu \mathrm{m}^3$, and were then loaded into a BeCu diamond anvil cell with a 400-$\mu$m-diameter culet. A 100-$\mu$m-thick pre-indented rhenium gasket was covered with boron nitride for electrical insulation and a 200-$\mu$m-diameter hole was drilled as the sample chamber. Daphne oil 7373 was used as the pressure transmitting medium, in order to obtain good hydrostaticity. The DAC was loaded together with several small ruby balls for pressure determination at room temperature using the ruby fluorescence method~\cite{Mao}. Electrical resistance measurements in both the piston-cylinder and diamond anvil cells were performed using a four-probe method, for which 15~$\mu$m diameter gold wires were attached to the samples using silver epoxy paste. The resistance was measured using a Teslatron-PT system with an Oxford $^{3}$He refrigerator and a Quantum Design Physical Property Measurement System (PPMS). Note that to avoid any extrinsic heating effects, for resistance measurements of samples S1-2 and S2-4, a current of 1 $\mu$A was used, while other samples were measured with a 100 $\mu$A current. The Hall resistivity was measured using a PPMS. Measurements of current–voltage ($I$–$V$) characteristics under pressure were performed in a Teslatron-PT system with an Oxford $^{3}$He refrigerator, where the current was generated by a Keithley 6221 unit and the voltage was measured with a Keithley 2182A nanovoltmeter. The chemical composition was checked using energy-dispersive x-ray analysis with a Hitachi SU-8010 field emission scanning electron microscope.
\clearpage

\hspace*{\fill}
\begin{figure}[h]
\centering
\includegraphics[width=0.95\columnwidth]{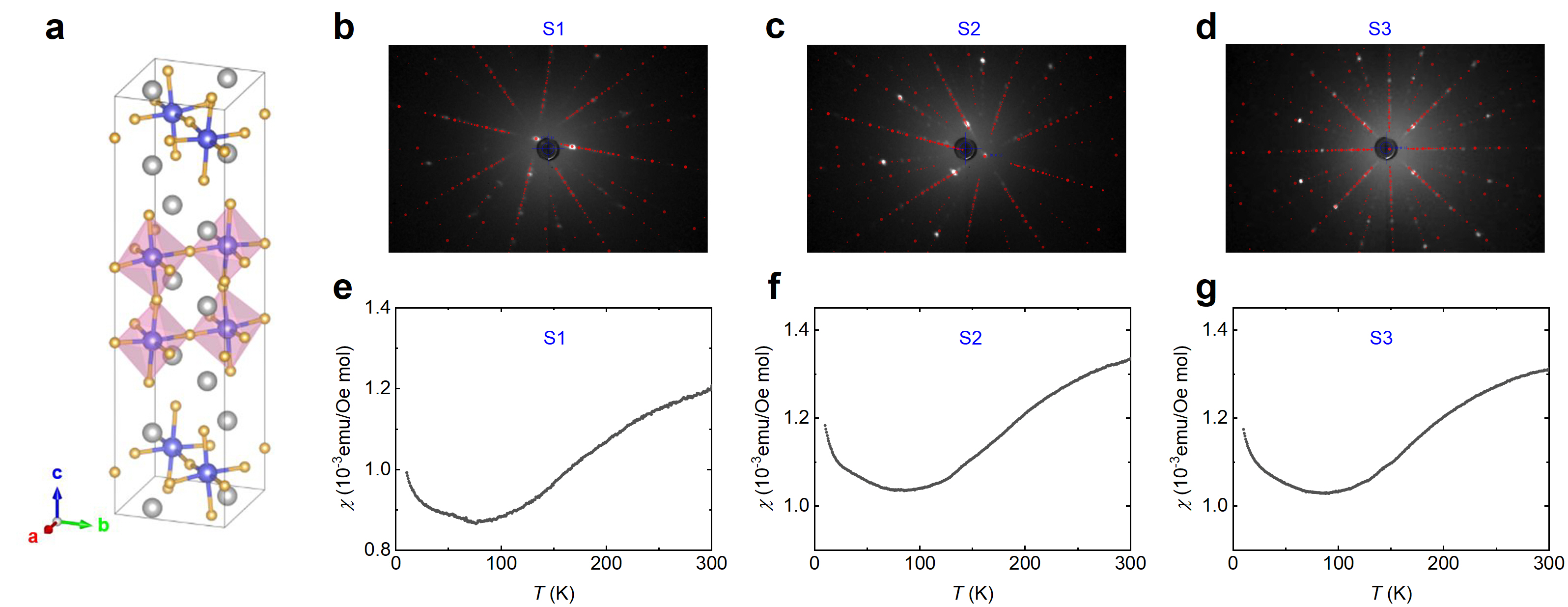}
\caption*{
\textbf{Extended Data Fig. 1 $|$ Crystal structure, Laue measurements, and magnetic susceptibility of single crystalline \LNO.} a, Crystal structure. b, c, d, X-ray Laue pattern corresponding to the [001] direction, where simulations of the orthorhombic structure with space group Amam are shown by red spots~\cite{Ekin}. e, f, g Temperature dependence of the magnetic susceptibility $\chi$($T$) of samples S1-S3 measured in a magnetic field of 0.4~T applied perpendicular to the $c$ axis. The temperature dependence of the magnetic susceptibility of the three samples are also displayed in this figure for fields perpendicular to the $c$-axis, which show similar behaviour, and also correspond well to the data in Ref.~\cite{Hundley2001PRB} for this field direction.
}
\label{FigE7}
\end{figure}

\begin{figure}
\centering
\includegraphics[width=0.6\columnwidth]{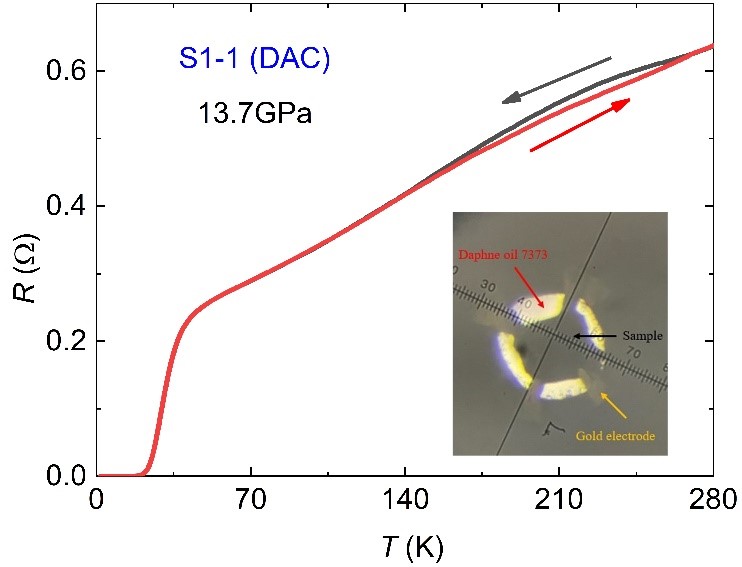}
\caption*{
\textbf{Extended Data Fig. 2 $|$ Resistance near the structural phase transition.} Resistance of \LNO\ sample S1-1 at 13.7~GPa as a function of temperature from 280~K down to 2~K. The black (red) arrow denotes the data taken upon cooling (warming). At high temperatures there is both a weak anomaly and hysteresis between warming and cooling, which might be associated with a structural phase transition. Note that in Ref.~\cite{Wang2023arxiv} a change of room temperature crystal structure is observed above 10~GPa, and therefore this anomaly may correspond to a transition between the corresponding structural phases. The inset shows the sample configuration in the DAC after pressure loading. The resistance was measured using a four-probe method. The yellow and red arrows point to the gold electrode and the pressure-transmitting medium Daphne oil 7373, respectively. The black region in the middle is the sample, and the rubies and silver epoxy paste are underneath the sample.
}
\label{FigE1}
\end{figure}

\hspace*{\fill}
\begin{figure}[h]
\centering
\includegraphics[width=0.95\columnwidth]{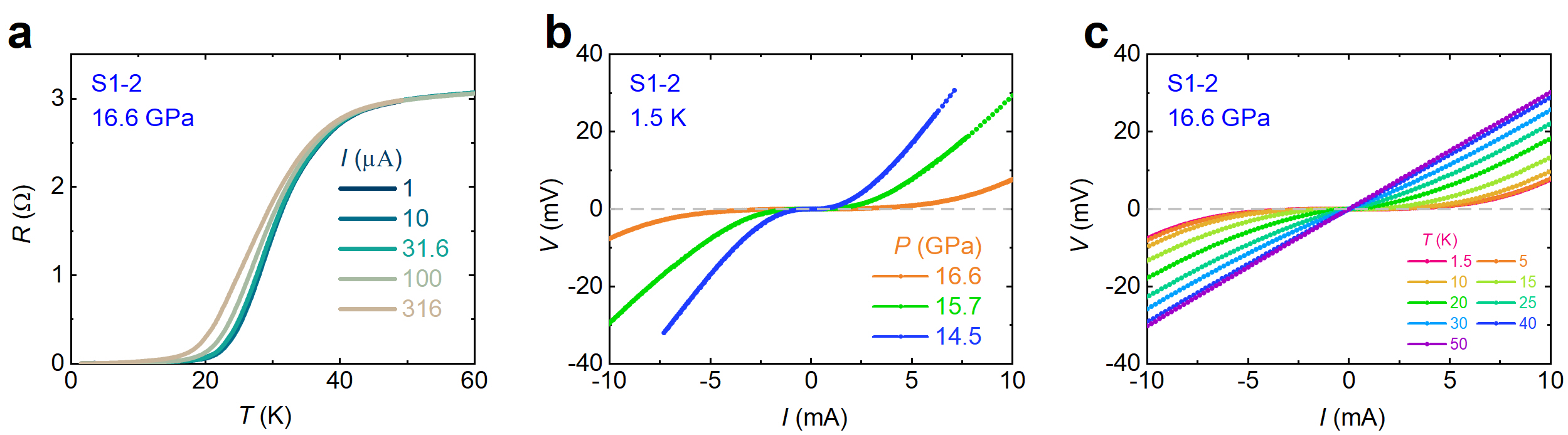}
\caption*{
\textbf{Extended Data Fig. 3 $|$ $R(T)$ curves at 16.6~GPa under different currents and $I$-$V$ curves at different pressures and temperatures.} a, Resistance of sample S1-2 at 16.6~GPa in the vicinity of the superconducting transition, measured with different excitation currents. For currents below 31.6~$\mu$A, the resistance curve remains unaffected by the current magnitude. b, $I$-$V$ curves of sample S1-2 at 1.5~K under various pressures. c, $I$-$V$ curves of sample S1-2 at various temperatures under a pressure of 16.6~GPa. The critical current at 1.5~K is around 3~mA, and the sample has a cross-sectional area of 70~$\mu$m $\times$ 5~$\mu$m, allowing us to estimate a critical current density of 850~A/cm$^2$.
}
\label{FigE2}
\end{figure}

\hspace*{\fill}
\begin{figure}[h]
\centering
\includegraphics[width=0.95\columnwidth]{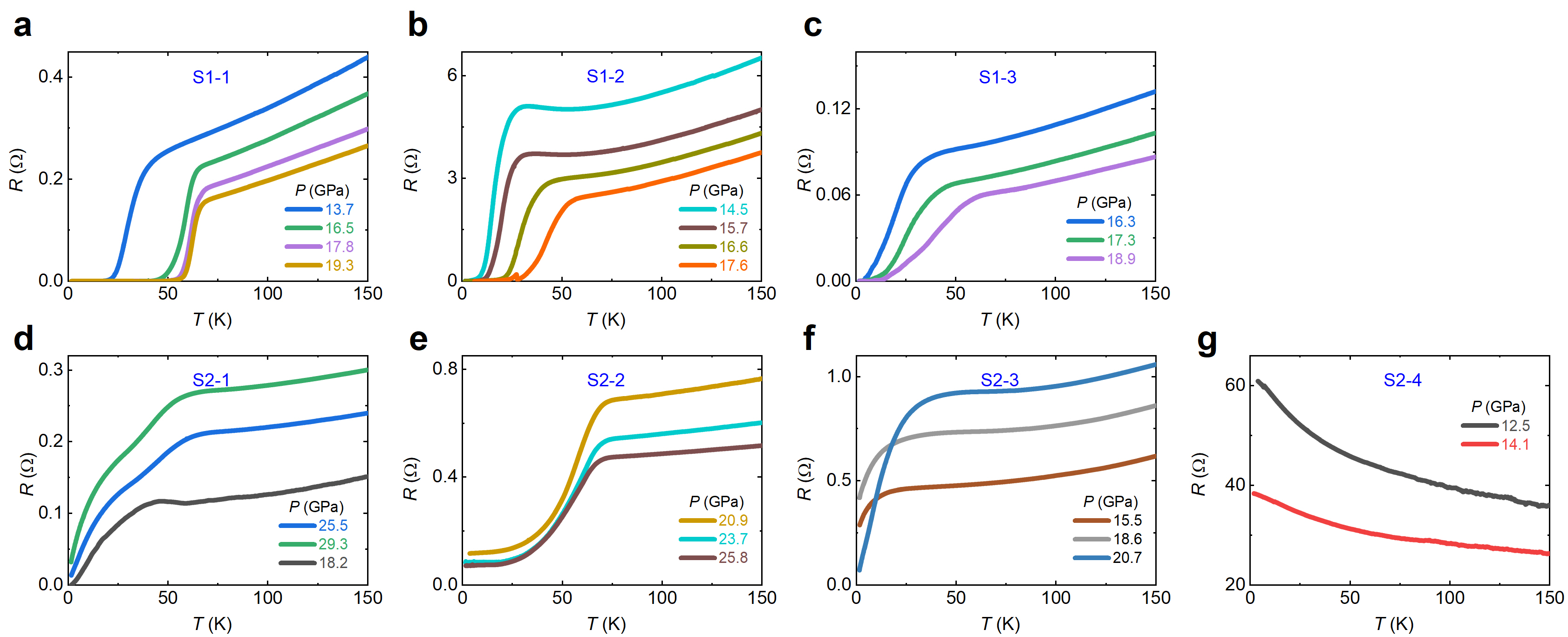}
\caption*{
\textbf{Extended Data Fig. 4 $|$ Resistance of \LNO\ single crystal samples under pressure.} We measured the $R(T)$ curves of seven \LNO\ single crystal samples under pressure. Four samples exhibited zero resistance at high pressures, and energy-dispersive x-ray analysis (EDX) measurements (Extended Data Fig. 4) show that three of these samples are more homogeneous and close to the stoichiometric composition (S1-1, S1-2, S1-3), while EDX of sample S2-1 was not measured as it broke upon releasing the pressure. For sample S1-3, in order to measure the Hall resistivity at high pressures (Extended Data Fig. 8), we used a 500-$\mu$m-diameter culet, which results in a more uneven pressure distribution near its maximum pressure of 20~GPa, leading to a broader superconducting transition at 18.9~GPa for this sample. The three samples (S2-2, S2-3, S2-4) with compositions far from 3:2:7 with large $\delta$ did not show zero resistance.
}
\label{FigE3}
\end{figure}

\hspace*{\fill}

\begin{figure}[h]
\centering
\includegraphics[width=0.95\columnwidth]{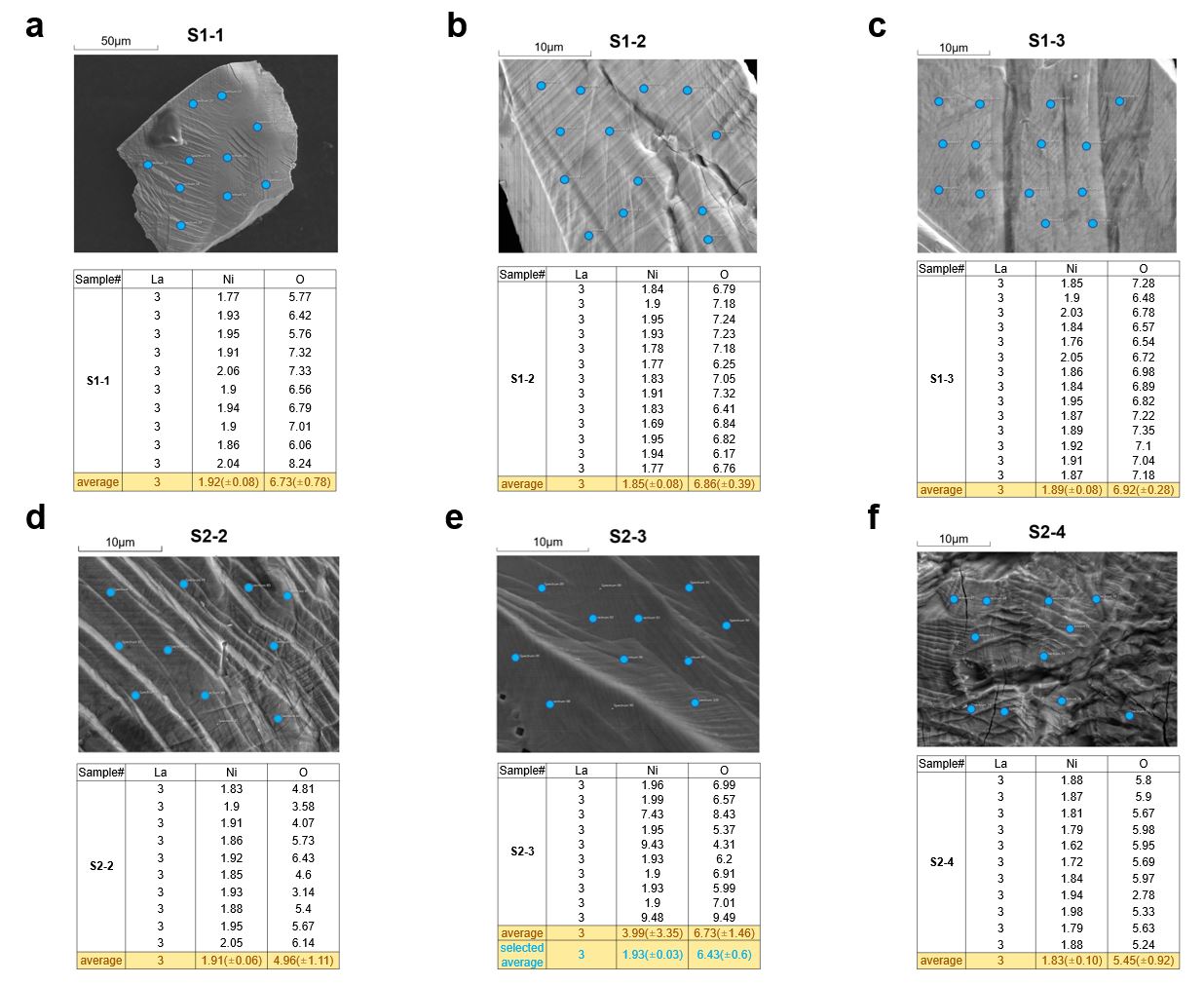}
\caption*{
\textbf{Extended Data Fig. 5 $|$ Scanning electron microscope images and chemical compositions of different \LNO\ single crystal samples.} Scanning electron microscope (SEM) images were taken of six of the small samples that were measured in a DAC. For each sample, we used EDX to check the chemical composition. We randomly selected various positions (marked with blue solid dots) for chemical composition analysis, where the La content was normalized to 3. The final results are listed in the tables below the SEM images, where the mean values are also given, as well as the standard deviation in parentheses. On average, the atomic ratios of La:Ni:O for samples S1-1, S1-2, and S1-3 are close to the stoichiometric values of 3:2:7, with smaller standard deviations indicating that those samples are more homogeneous. However, the atomic ratios of S2-2, S2-3, and S2-4 deviate more significantly from stoichiometry, with larger variations in the compositions between different points across the sample. For S2-3, we also include a “selected average” where the regions with Ni $\gg$ 2 are excluded from the statistics.
}
\label{FigE4}
\end{figure}

\clearpage
\begin{figure}
\centering
\includegraphics[width=0.7\columnwidth]{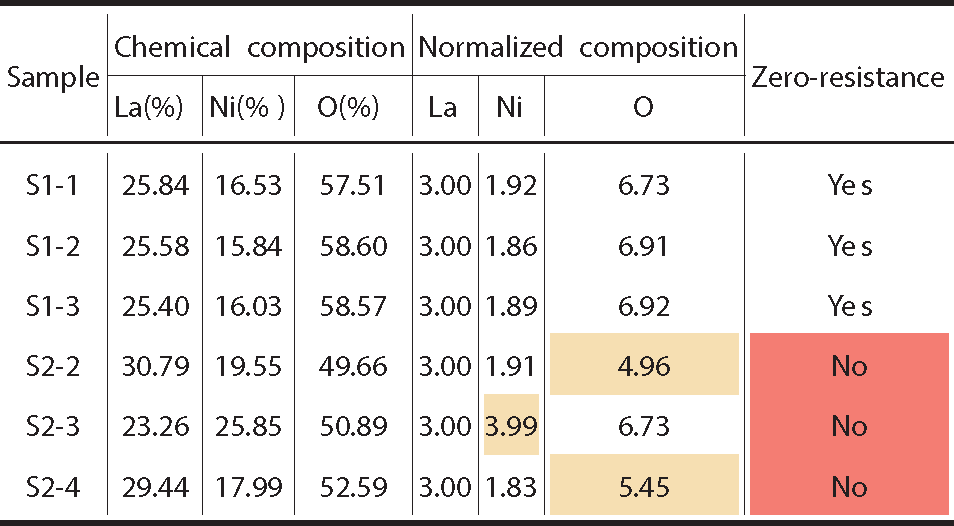}
\caption*{
\textbf{Extended Data Table 1 $|$ Summary of the comparison between EDX and electrical resistance measurements.} The yellow highlights indicate a significant deviation from the stoichiometric composition, while the red highlights show the samples without zero-resistance.
}
\end{figure}
\hspace*{\fill}

\clearpage
\begin{figure}[h]
\centering
\includegraphics[width=0.95\columnwidth]{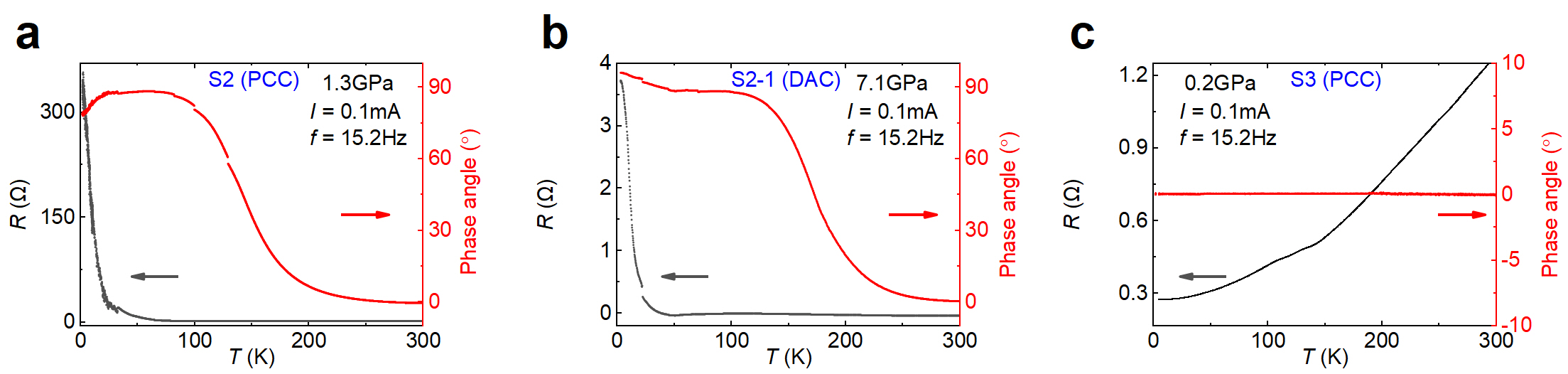}
\caption*{
\textbf{Extended Data Fig. 6 $|$ Possible origin of the extrinsic insulating behaviour.} We conducted measurements on samples S2 and S2-1 which were subjected to thermal treatment (heated for 2~hours at 400~K) and sample S3 which had no thermal treatment. After the thermal treatment, as shown in a and b, the $R(T)$ curves (black left axis) at 1.2~GPa and 7.1~GPa both exhibit weakly insulating behaviour. However, it can be seen that the phase angle (red right axis) gradually deviates from zero with decreasing temperature, indicating that the measured resistance is not the intrinsic signal. At the same time, we found that the contact resistance between the silver epoxy paste and sample increased gradually from a few ohms at 300~K, to several tens of kilo-ohms at 2~K, which explains the significant deviation in the phase angle. In contrast, for the sample without heat treatment the contact resistance between the sample and silver epoxy paste remains at the level of ohms and does not change with temperature. As shown in c, as the temperature decreases the phase angle remains nearly zero, and the $R(T)$ curve exhibits metallic behaviour. This suggests that the additional heat treatment may lead to oxygen vacancies, resulting in the formation of an insulating layer on the surface and an increase of the contact resistance upon cooling. This phenomenon has also been observed in YBCO~\cite{Blankenship1988YBCO}.
}
\label{FigE5}
\end{figure}

\hspace*{\fill}

\begin{figure}[h]
\centering
\includegraphics[width=0.95\columnwidth]{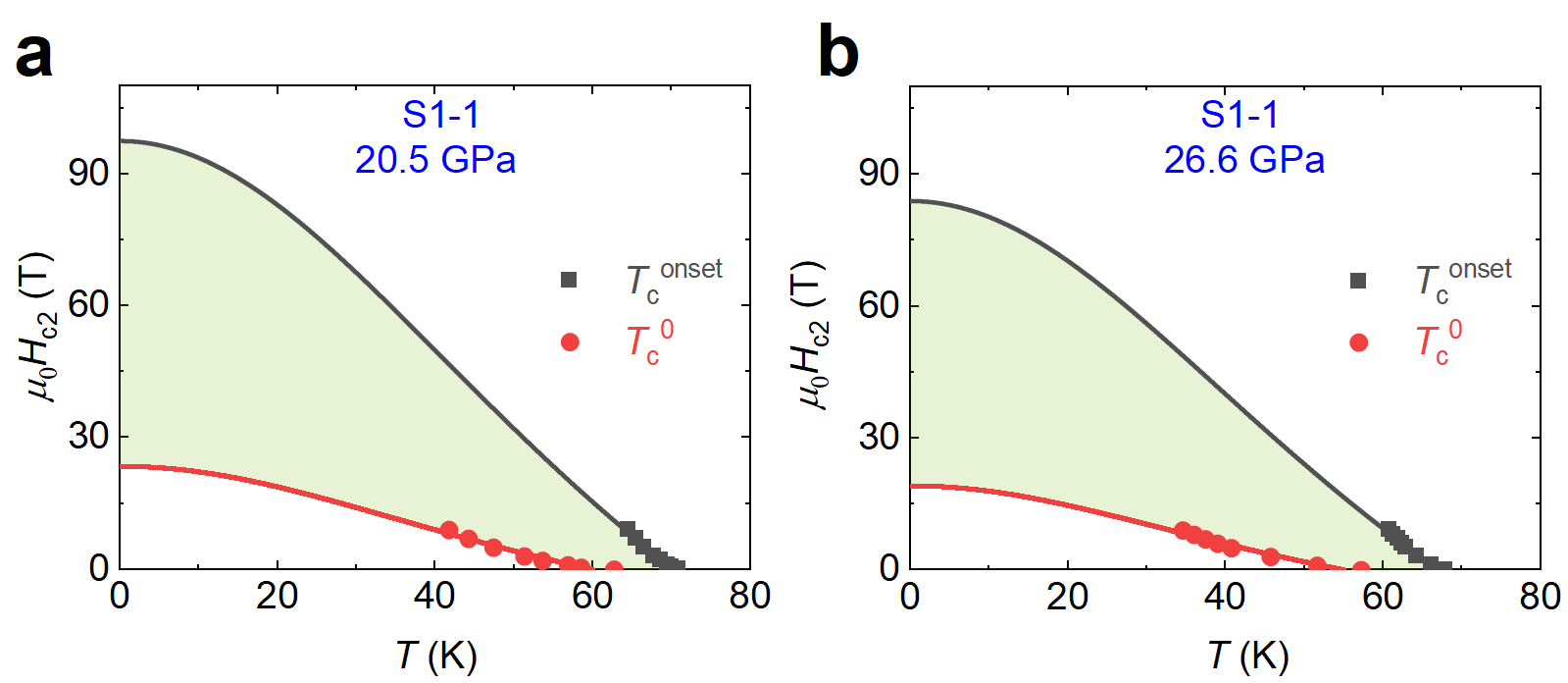}
\caption*{
\textbf{Extended Data Fig. 7 $|$ Temperature dependence of the upper critical fields at two pressures.} The red and black points in both panels represent the upper critical fields determined by $T_c^0$ and $T_c^{onset}$, respectively, which are determined from the construction shown in Fig. 1. The solid lines show the results of fitting with a Ginzburg-Landau model, and the green region between red and black solid lines represents the region where there is an intermediate state or a thermally activated flux flow. By fitting $H_{c2}$($T$) determined from $T_c^0$ with a Ginzburg-Landau model, we obtained $H_{c2}$(0) of 23~T and 19~T for 20.5~GPa and 26.6~GPa, respectively. By fitting $H_{c2}$($T$) determined from $T_c^{onset}$, respective zero-temperature values of 97~T and 83~T are obtained.
}
\label{FigE6}
\end{figure}
\hspace*{\fill}

\begin{figure}[h]
\centering
\includegraphics[width=0.95\columnwidth]{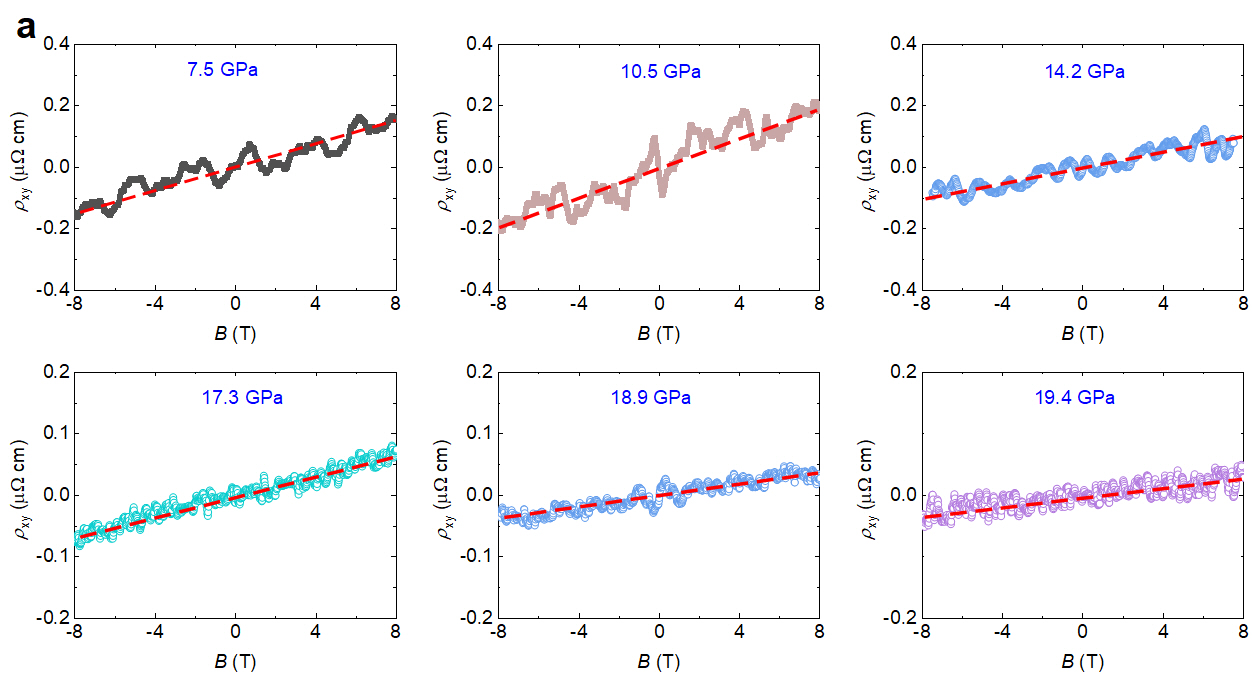}
\caption*{
\textbf{Extended Data Fig. 8 $|$ Hall resistivity $\rho_{xy}$($H$) at 80~K under various pressures.} $\rho_{xy}$($H$) for \LNO\ sample S1-3 from 7.5~GPa to 19.4~GPa, where the red dashed lines show fits to a linear magnetic field dependence.
}
\label{FigE8}
\end{figure}


\clearpage
\begin{thebibliography}{10}

\bibitem{Wang2023arxiv}
\bibinfo{author}{Sun, H.} \emph{et~al.}
\newblock \bibinfo{title}{Signatures of superconductivity near
  80{\hspace{0.167em}}{K} in a nickelate under high pressure}.
\newblock \emph{\bibinfo{journal}{Nature}} \textbf{\bibinfo{volume}{621}},
  \bibinfo{pages}{493--498} (\bibinfo{year}{2023}).

\bibitem{Cooper2009}
\bibinfo{author}{Cooper, R.~A.} \emph{et~al.}
\newblock \bibinfo{title}{Anomalous {Criticality} in the {Electrical}
  {Resistivity} of {La$_{2\ensuremath{-}{x}}$Sr$_x$CuO$_4$}}.
\newblock \emph{\bibinfo{journal}{Science}} \textbf{\bibinfo{volume}{323}},
  \bibinfo{pages}{603--607} (\bibinfo{year}{2009}).

\bibitem{Taillefer2010}
\bibinfo{author}{Taillefer, L.}
\newblock \bibinfo{title}{Scattering and pairing in cuprate superconductors}.
\newblock \emph{\bibinfo{journal}{Annu. Rev. Condens. Matter Phys.}}
  \textbf{\bibinfo{volume}{1}}, \bibinfo{pages}{51--70} (\bibinfo{year}{2010}).

\bibitem{Jiang2023NP}
\bibinfo{author}{Jiang, X.} \emph{et~al.}
\newblock \bibinfo{title}{Interplay between superconductivity and the
  strange-metal state in {FeSe}}.
\newblock \emph{\bibinfo{journal}{Nat. Phys.}} \textbf{\bibinfo{volume}{19}},
  \bibinfo{pages}{365--371} (\bibinfo{year}{2023}).

\bibitem{Greene2020}
\bibinfo{author}{Greene, R.~L.}, \bibinfo{author}{Mandal, P.~R.},
  \bibinfo{author}{Poniatowski, N.~R.} \& \bibinfo{author}{Sarkar, T.}
\newblock \bibinfo{title}{The {Strange} {Metal} {State} of the
  {Electron}-{Doped} {Cuprates}}.
\newblock \emph{\bibinfo{journal}{Annual Review of Condensed Matter Physics}}
  \textbf{\bibinfo{volume}{11}}, \bibinfo{pages}{213--229}
  (\bibinfo{year}{2020}).
\newblock

\bibitem{Phillips2022}
\bibinfo{author}{Phillips, P.~W.}, \bibinfo{author}{Hussey, N.~E.} \&
  \bibinfo{author}{Abbamonte, P.}
\newblock \bibinfo{title}{Stranger than metals}.
\newblock \emph{\bibinfo{journal}{Science}} \textbf{\bibinfo{volume}{377}}
  (\bibinfo{year}{2022}).


\expandafter\ifx\csname url\endcsname\relax
  \def\url#1{\texttt{#1}}\fi
\expandafter\ifx\csname urlprefix\endcsname\relax\def\urlprefix{URL }\fi
\providecommand{\bibinfo}[2]{#2}
\providecommand{\eprint}[2][]{\url{#2}}

\bibitem{Norman2014}
\bibinfo{author}{Norman, M.~R.}
\newblock \emph{\bibinfo{title}{Unconventional {Superconductivity}}}.
\newblock Novel superfluids (\bibinfo{publisher}{Oxford University Press},
  \bibinfo{address}{Oxford, UK}, \bibinfo{year}{2014}).

\bibitem{stewart2017}
\bibinfo{author}{Stewart, G.~R.}
\newblock \bibinfo{title}{Unconventional superconductivity}.
\newblock \emph{\bibinfo{journal}{Advances in Physics}}
  \textbf{\bibinfo{volume}{66}}, \bibinfo{pages}{75--196}
  (\bibinfo{year}{2017}).
\newblock \urlprefix\url{https://doi.org/10.1080/00018732.2017.1331615}.

\bibitem{Muller1986}
\bibinfo{author}{Bednorz, J.~G.} \& \bibinfo{author}{M{\"u}ller, K.~A.}
\newblock \bibinfo{title}{Possible high {$T{\rm _c}$} superconductivity in the
  {Ba-La-Cu-O} system}.
\newblock \emph{\bibinfo{journal}{Zeitschrift f\"ur Physik B Condensed Matter}}
  \textbf{\bibinfo{volume}{64}}, \bibinfo{pages}{189--193}
  (\bibinfo{year}{1986}).
\newblock \urlprefix\url{https://doi.org/10.1007/bf01303701}.

\bibitem{Chu1987}
\bibinfo{author}{Chu, C.~W.} \emph{et~al.}
\newblock \bibinfo{title}{Evidence for superconductivity above 40~{K} in the
  {La-Ba-Cu-O} compound system}.
\newblock \emph{\bibinfo{journal}{Phys. Rev. Lett.}}
  \textbf{\bibinfo{volume}{58}}, \bibinfo{pages}{405--407}
  (\bibinfo{year}{1987}).
\newblock \urlprefix\url{https://link.aps.org/doi/10.1103/PhysRevLett.58.405}.

\bibitem{Hosono2008}
\bibinfo{author}{Kamihara, Y.}, \bibinfo{author}{Watanabe, T.},
  \bibinfo{author}{Hirano, M.} \& \bibinfo{author}{Hosono, H.}
\newblock \bibinfo{title}{Iron-{Based} {Layered} {Superconductor}
  {La(O$_{1\ensuremath{-}{x}}$F$_x$)FeAs} (x = 0.05-0.12) with {$T{\rm _c}$} =
  26~{K}}.
\newblock \emph{\bibinfo{journal}{Journal of the American Chemical Society}}
  \textbf{\bibinfo{volume}{130}}, \bibinfo{pages}{3296--3297}
  (\bibinfo{year}{2008}).
\newblock \urlprefix\url{https://doi.org/10.1021/ja800073m}.

\bibitem{Chen2008}
\bibinfo{author}{Chen, X.~H.} \emph{et~al.}
\newblock \bibinfo{title}{Superconductivity at 43 {{K}} in
  {SmFeAsO$_{1\ensuremath{-}{x}}$F$_x$}}.
\newblock \emph{\bibinfo{journal}{Nature}} \textbf{\bibinfo{volume}{453}},
  \bibinfo{pages}{761--762} (\bibinfo{year}{2008}).
\newblock \urlprefix\url{https://doi.org/10.1038/nature07045}.


\bibitem{Li2019nature}
\bibinfo{author}{Li, D.} \emph{et~al.}
\newblock \bibinfo{title}{Superconductivity in an infinite-layer nickelate}.
\newblock \emph{\bibinfo{journal}{Nature}} \textbf{\bibinfo{volume}{572}},
  \bibinfo{pages}{624--627} (\bibinfo{year}{2019}).
\newblock \urlprefix\url{https://doi.org/10.1038/s41586-019-1496-5}.

\bibitem{Ariando2020PRL}
\bibinfo{author}{Zeng, S.} \emph{et~al.}
\newblock \bibinfo{title}{Phase {Diagram} and {Superconducting} {Dome} of
  {Infinite}-{Layer} {Nd$_{1\ensuremath{-}x}$Sr$_x$NiO$_2$} {Thin} {Films}}.
\newblock \emph{\bibinfo{journal}{Phys. Rev. Lett.}}
  \textbf{\bibinfo{volume}{125}}, \bibinfo{pages}{147003}
  (\bibinfo{year}{2020}).
\newblock
  \urlprefix\url{https://link.aps.org/doi/10.1103/PhysRevLett.125.147003}.

\bibitem{Harold2020PRM}
\bibinfo{author}{Osada, M.}, \bibinfo{author}{Wang, B.~Y.},
  \bibinfo{author}{Lee, K.}, \bibinfo{author}{Li, D.} \&
  \bibinfo{author}{Hwang, H.~Y.}
\newblock \bibinfo{title}{Phase diagram of infinite layer praseodymium
  nickelate {Pr$_{1\ensuremath{-}x}$Sr$_x$NiO$_2$} thin films}.
\newblock \emph{\bibinfo{journal}{Phys. Rev. Mater.}}
  \textbf{\bibinfo{volume}{4}}, \bibinfo{pages}{121801} (\bibinfo{year}{2020}).
\newblock
  \urlprefix\url{https://link.aps.org/doi/10.1103/PhysRevMaterials.4.121801}.

\bibitem{Ariando2022scienceadvance}
\bibinfo{author}{Zeng, S.} \emph{et~al.}
\newblock \bibinfo{title}{Superconductivity in infinite-layer nickelate
  {La$_{1\ensuremath{-}x}$Ca$_x$NiO$_2$} thin films}.
\newblock \emph{\bibinfo{journal}{Sci. Adv.}} \textbf{\bibinfo{volume}{8}},
  \bibinfo{pages}{eabl9927} (\bibinfo{year}{2022}).
\newblock
  \urlprefix\url{https://www.science.org/doi/abs/10.1126/sciadv.abl9927}.

\bibitem{Ding2023nature}
\bibinfo{author}{Ding, X.} \emph{et~al.}
\newblock \bibinfo{title}{Critical role of hydrogen for superconductivity in
  nickelates}.
\newblock \emph{\bibinfo{journal}{Nature}} \textbf{\bibinfo{volume}{615}},
  \bibinfo{pages}{50--55} (\bibinfo{year}{2023}).
\newblock \urlprefix\url{https://doi.org/10.1038/s41586-022-05657-2}.

\bibitem{JGCheng2022NC}
\bibinfo{author}{Wang, N.~N.} \emph{et~al.}
\newblock \bibinfo{title}{Pressure induced monotonic enhancement of {$T{\rm
  _c}$} to over 30~{K} in superconducting {Pr$_{0.82}$Sr$_{0.18}$NiO$_2$}}.
\newblock \emph{\bibinfo{journal}{Nat. Commun.}} \textbf{\bibinfo{volume}{13}}
  (\bibinfo{year}{2022}).
\newblock \urlprefix\url{https://doi.org/10.1038/s41467-022-32065-x}.

\bibitem{WenHaiHu2020CM}
\bibinfo{author}{Li, Q.} \emph{et~al.}
\newblock \bibinfo{title}{Absence of superconductivity in bulk
  {Nd$_{1\ensuremath{-}x}$Sr$_x$NiO$_2$}}.
\newblock \emph{\bibinfo{journal}{Commun. Mater.}} \textbf{\bibinfo{volume}{1}}
  (\bibinfo{year}{2020}).
\newblock \urlprefix\url{https://doi.org/10.1038/s43246-020-0018-1}.

\bibitem{Phelan2020PRM}
\bibinfo{author}{Wang, B.-X.} \emph{et~al.}
\newblock \bibinfo{title}{Synthesis and characterization of bulk
  {Nd$_{1\ensuremath{-}x}$Sr$_x$NiO$_2$} and
  {Nd$_{1\ensuremath{-}x}$Sr$_x$NiO$_3$}}.
\newblock \emph{\bibinfo{journal}{Phys. Rev. Mater.}}
  \textbf{\bibinfo{volume}{4}}, \bibinfo{pages}{084409} (\bibinfo{year}{2020}).
\newblock
  \urlprefix\url{https://link.aps.org/doi/10.1103/PhysRevMaterials.4.084409}.

\bibitem{Nakata}
\bibinfo{author}{Nakata, M.}, \bibinfo{author}{Ogura, D.},
  \bibinfo{author}{Usui, H.} \& \bibinfo{author}{Kuroki, K.}
\newblock \bibinfo{title}{Finite-energy spin fluctuations as a pairing glue in
  systems with coexisting electron and hole bands}.
\newblock \emph{\bibinfo{journal}{Phys. Rev. B}} \textbf{\bibinfo{volume}{95}},
  \bibinfo{pages}{214509} (\bibinfo{year}{2017}).
\newblock \urlprefix\url{https://link.aps.org/doi/10.1103/PhysRevB.95.214509}.

\bibitem{YaoDaoXin2023}
\bibinfo{author}{Luo, Z.}, \bibinfo{author}{Hu, X.}, \bibinfo{author}{Wang,
  M.}, \bibinfo{author}{W\'u, W.} \& \bibinfo{author}{Yao, D.-X.}
\newblock \bibinfo{title}{Bilayer two-orbital model of
  {$\mathrm{L}{\mathrm{a}}_{3}\mathrm{N}{\mathrm{i}}_{2}{\mathrm{O}}_{7}$}
  under pressure}.
\newblock \emph{\bibinfo{journal}{Phys. Rev. Lett.}}
  \textbf{\bibinfo{volume}{131}}, \bibinfo{pages}{126001}
  (\bibinfo{year}{2023}).
\newblock
  \urlprefix\url{https://link.aps.org/doi/10.1103/PhysRevLett.131.126001}.

\bibitem{Elbio2023ARXIV}
\bibinfo{author}{Zhang, Y.}, \bibinfo{author}{Lin, L.-F.},
  \bibinfo{author}{Moreo, A.} \& \bibinfo{author}{Dagotto, E.}
\newblock \bibinfo{title}{Electronic structure, dimer physics,
  orbital-selective behavior, and magnetic tendencies in the bilayer nickelate
  superconductor la$_3$ni$_2$o$_7$ under pressure}.
\newblock \emph{\bibinfo{journal}{Phys. Rev. B}}
  \textbf{\bibinfo{volume}{108}}, \bibinfo{pages}{L180510}
  (\bibinfo{year}{2023}).
\newblock
  \urlprefix\url{https://link.aps.org/doi/10.1103/PhysRevB.108.L180510}.

\bibitem{HuJiangping2023ARXIV}
\bibinfo{author}{Gu, Y.}, \bibinfo{author}{Le, C.}, \bibinfo{author}{Yang, Z.},
  \bibinfo{author}{Wu, X.} \& \bibinfo{author}{Hu, J.}
\newblock \bibinfo{title}{Effective model and pairing tendency in bilayer
  {Ni}-based superconductor {La$_3$Ni$_2$O$_7$}} (\bibinfo{year}{2023}).
\newblock \urlprefix\url{https://arxiv.org/abs/2306.07275}.

\bibitem{Eremin2023ARXIV}
\bibinfo{author}{Lechermann, F.}, \bibinfo{author}{Gondolf, J.},
  \bibinfo{author}{B\"{o}tzel, S.} \& \bibinfo{author}{Eremin, I.~M.}
\newblock \bibinfo{title}{Electronic correlations and superconducting
  instability in la$_3$ni$_2$o$_7$ under high pressure}.
\newblock \emph{\bibinfo{journal}{Phys. Rev. B}}
  \textbf{\bibinfo{volume}{108}}, \bibinfo{pages}{L201121}
  (\bibinfo{year}{2023}).
\newblock
  \urlprefix\url{https://link.aps.org/doi/10.1103/PhysRevB.108.L201121}.

\bibitem{WANGQIANGHUA2023ARXIV}
\bibinfo{author}{Yang, Q.-G.}, \bibinfo{author}{Wang, D.} \&
  \bibinfo{author}{Wang, Q.-H.}
\newblock \bibinfo{title}{Possible ${s}_{\ifmmode\pm\else\textpm\fi{}}$-wave
  superconductivity in ${\mathrm{la}}_{3}{\mathrm{ni}}_{2}{\mathrm{o}}_{7}$}.
\newblock \emph{\bibinfo{journal}{Phys. Rev. B}}
  \textbf{\bibinfo{volume}{108}}, \bibinfo{pages}{L140505}
  (\bibinfo{year}{2023}).
\newblock
  \urlprefix\url{https://link.aps.org/doi/10.1103/PhysRevB.108.L140505}.

\bibitem{Kazuhiko2023ARXIV}
\bibinfo{author}{Sakakibara, H.}, \bibinfo{author}{Kitamine, N.},
  \bibinfo{author}{Ochi, M.} \& \bibinfo{author}{Kuroki, K.}
\newblock \bibinfo{title}{Possible high ${T}_{c}$ superconductivity in
  ${\mathrm{la}}_{3}{\mathrm{ni}}_{2}{\mathrm{o}}_{7}$ under high pressure
  through manifestation of a nearly half-filled bilayer hubbard model}.
\newblock \emph{\bibinfo{journal}{Phys. Rev. Lett.}}
  \textbf{\bibinfo{volume}{132}}, \bibinfo{pages}{106002}
  (\bibinfo{year}{2024}).
\newblock
  \urlprefix\url{https://link.aps.org/doi/10.1103/PhysRevLett.132.106002}.

\bibitem{ZhangGuangMing2023ARXIV}
\bibinfo{author}{Shen, Y.}, \bibinfo{author}{Qin, M.} \&
  \bibinfo{author}{Zhang, G.-M.}
\newblock \bibinfo{title}{Effective bi-layer model hamiltonian and
  density-matrix renormalization group study for the high-tc superconductivity
  in{La$_3$Ni$_2$O$_7$} under high pressure}.
\newblock \emph{\bibinfo{journal}{Chinese Physics Letters}}
  \textbf{\bibinfo{volume}{40}}, \bibinfo{pages}{127401}
  (\bibinfo{year}{2023}).
\newblock \urlprefix\url{https://dx.doi.org/10.1088/0256-307X/40/12/127401}.

\bibitem{LeonovPRB2023}
\bibinfo{author}{Shilenko, D.~A.} \& \bibinfo{author}{Leonov, I.~V.}
\newblock \bibinfo{title}{Correlated electronic structure, orbital-selective
  behavior, and magnetic correlations in double-layer {La$_3$Ni$_2$O$_7$} under
  pressure}.
\newblock \emph{\bibinfo{journal}{Phys. Rev. B}}
  \textbf{\bibinfo{volume}{108}}, \bibinfo{pages}{125105}
  (\bibinfo{year}{2023}).
\newblock \urlprefix\url{https://link.aps.org/doi/10.1103/PhysRevB.108.125105}.

\bibitem{WenHaiHuARXIV2023}
\bibinfo{author}{Liu, Z.} \emph{et~al.}
\newblock \bibinfo{title}{Electronic correlations and energy gap in the bilayer
  nickelate {La$_3$Ni$_2$O$_7$}} (\bibinfo{year}{2023}).
\newblock \urlprefix\url{https://arxiv.org/abs/2307.02950}.

\bibitem{Philipp2023ARXIV}
\bibinfo{author}{Christiansson, V.}, \bibinfo{author}{Petocchi, F.} \&
  \bibinfo{author}{Werner, P.}
\newblock \bibinfo{title}{Correlated electronic structure of
  {La$_3$Ni$_2$O$_7$} under pressure}.
\newblock \emph{\bibinfo{journal}{Phys. Rev. Lett.}}
  \textbf{\bibinfo{volume}{131}}, \bibinfo{pages}{206501}
  (\bibinfo{year}{2023}).
\newblock
  \urlprefix\url{https://link.aps.org/doi/10.1103/PhysRevLett.131.206501}.

\bibitem{YangYifengARXIV2023}
\bibinfo{author}{Cao, Y.} \& \bibinfo{author}{Yang, Y.-F.}
\newblock \bibinfo{title}{Flat bands promoted by hund's rule coupling in the
  candidate double-layer high-temperature superconductor {La$_3$Ni$_2$O$_7$}
  under high pressure}.
\newblock \emph{\bibinfo{journal}{Phys. Rev. B}}
  \textbf{\bibinfo{volume}{109}}, \bibinfo{pages}{L081105}
  (\bibinfo{year}{2024}).
\newblock
  \urlprefix\url{https://link.aps.org/doi/10.1103/PhysRevB.109.L081105}.

\bibitem{CHEN2023TEM}
\bibinfo{author}{Dong, Z.} \emph{et~al.}
\newblock \bibinfo{title}{Visualization of oxygen vacancies and self-doped
  ligand holes in {La$_3$Ni$_2$O$_{7-\delta}$}} (\bibinfo{year}{2023}).
\newblock \urlprefix\url{https://arxiv.org/abs/2312.15727}.

\bibitem{Wang2022sciencechina}
\bibinfo{author}{Liu, Z.} \emph{et~al.}
\newblock \bibinfo{title}{Evidence for charge and spin density waves in single
  crystals of {La$_3$Ni$_2$O$_7$} and {La$_3$Ni$_2$O$_6$}}.
\newblock \emph{\bibinfo{journal}{Sci. China-Phys. Mech. Astron.}}
  \textbf{\bibinfo{volume}{66}}, \bibinfo{pages}{217411}
  (\bibinfo{year}{2023}).
\newblock \urlprefix\url{https://doi.org/10.1007/s11433-022-1962-4}.

\bibitem{Hundley2001PRB}
\bibinfo{author}{Wu, G.}, \bibinfo{author}{Neumeier, J.~J.} \&
  \bibinfo{author}{Hundley, M.~F.}
\newblock \bibinfo{title}{Magnetic susceptibility, heat capacity, and pressure
  dependence of the electrical resistivity of {La$_3$Ni$_2$O$_7$} and
  {La$_4$Ni$_3$O$_{10}$}}.
\newblock \emph{\bibinfo{journal}{Phys. Rev. B}} \textbf{\bibinfo{volume}{63}},
  \bibinfo{pages}{245120} (\bibinfo{year}{2001}).
\newblock \urlprefix\url{https://link.aps.org/doi/10.1103/PhysRevB.63.245120}.

\bibitem{ZHANG1994}
\bibinfo{author}{Zhang, Z.}, \bibinfo{author}{Greenblatt, M.} \&
  \bibinfo{author}{Goodenough, J.}
\newblock \bibinfo{title}{Synthesis, {Structure}, and {Properties} of the
  {Layered} {Perovskite} {La$_{3}$Ni$_{2}$O$_{7\ensuremath{-}{\delta}}$}}.
\newblock \emph{\bibinfo{journal}{J. Solid State Chem.}}
  \textbf{\bibinfo{volume}{108}}, \bibinfo{pages}{402--409}
  (\bibinfo{year}{1994}).
\newblock
  \urlprefix\url{https://www.sciencedirect.com/science/article/pii/S0022459684710590}.

\bibitem{Masatoshi1995JPSJ}
\bibinfo{author}{Taniguchi, S.} \emph{et~al.}
\newblock \bibinfo{title}{Transport, Magnetic and Thermal Properties of
  La$_{3}$Ni$_{2}$O$_{7\ensuremath{-}{\delta}}$}.
\newblock \emph{\bibinfo{journal}{J. Phys. Soc. Jpn.}}
  \textbf{\bibinfo{volume}{64}}, \bibinfo{pages}{1644--1650}
  (\bibinfo{year}{1995}).
\newblock \urlprefix\url{https://doi.org/10.1143/JPSJ.64.1644}.

\bibitem{Blankenship1988YBCO}
\bibinfo{author}{Ekin, J.~W.} \emph{et~al.}
\newblock \bibinfo{title}{{High $T_{\rm c}$ superconductor$/$noble‐metal
  contacts with surface resistivities in the $10^{-10}\Omega\ \mathrm{cm}^2$
  range}}.
\newblock \emph{\bibinfo{journal}{Appl. Phys. Lett.}}
  \textbf{\bibinfo{volume}{52}}, \bibinfo{pages}{1819--1821}
  (\bibinfo{year}{1988}).
\newblock \urlprefix\url{https://doi.org/10.1063/1.99725}.

\bibitem{Vinokur1994RMP}
\bibinfo{author}{Blatter, G.}, \bibinfo{author}{Feigel'man, M.~V.},
  \bibinfo{author}{Geshkenbein, V.~B.}, \bibinfo{author}{Larkin, A.~I.} \&
  \bibinfo{author}{Vinokur, V.~M.}
\newblock \bibinfo{title}{Vortices in high-temperature superconductors}.
\newblock \emph{\bibinfo{journal}{Rev. Mod. Phys.}}
  \textbf{\bibinfo{volume}{66}}, \bibinfo{pages}{1125--1388}
  (\bibinfo{year}{1994}).
\newblock \urlprefix\url{https://link.aps.org/doi/10.1103/RevModPhys.66.1125}.

\bibitem{Wen2008}
\bibinfo{author}{Jaroszynski, J.} \emph{et~al.}
\newblock \bibinfo{title}{Upper critical fields and thermally-activated
  transport of {NdFeAsO$_{0.7}$F$_{0.3}$} single crystal}.
\newblock \emph{\bibinfo{journal}{Phys. Rev. B}} \textbf{\bibinfo{volume}{78}},
  \bibinfo{pages}{174523} (\bibinfo{year}{2008}).
\newblock \urlprefix\url{https://link.aps.org/doi/10.1103/PhysRevB.78.174523}.

\bibitem{Jiao2012}
\bibinfo{author}{Jiao, L.} \emph{et~al.}
\newblock \bibinfo{title}{Upper critical field and thermally activated flux
  flow in single-crystalline {Tl$_{0.58}$Rb${_{0.42}}$Fe$_{1.72}$Se$_2$}}.
\newblock \emph{\bibinfo{journal}{Phys. Rev. B}} \textbf{\bibinfo{volume}{85}},
  \bibinfo{pages}{064513} (\bibinfo{year}{2012}).
\newblock \urlprefix\url{https://link.aps.org/doi/10.1103/PhysRevB.85.064513}.

\bibitem{Waszczak1990PRB}
\bibinfo{author}{Palstra, T. T.~M.}, \bibinfo{author}{Batlogg, B.},
  \bibinfo{author}{van Dover, R.~B.}, \bibinfo{author}{Schneemeyer, L.~F.} \&
  \bibinfo{author}{Waszczak, J.~V.}
\newblock \bibinfo{title}{Dissipative flux motion in high-temperature
  superconductors}.
\newblock \emph{\bibinfo{journal}{Phys. Rev. B}} \textbf{\bibinfo{volume}{41}},
  \bibinfo{pages}{6621--6632} (\bibinfo{year}{1990}).
\newblock \urlprefix\url{https://link.aps.org/doi/10.1103/PhysRevB.41.6621}.

\bibitem{Ling2000}
\bibinfo{author}{Ling, C.~D.}, \bibinfo{author}{Argyriou, D.~N.},
  \bibinfo{author}{Wu, G.} \& \bibinfo{author}{Neumeier, J.}
\newblock \bibinfo{title}{Neutron diffraction study of {La$_3$Ni$_2$O$_7$:}
  {S}tructural relationships among $n=1$, 2, and 3 phases
  {La$_{n+1}$Ni$_n$O$_{3n+1}$}}.
\newblock \emph{\bibinfo{journal}{J. Solid State Chem}}
  \textbf{\bibinfo{volume}{152}}, \bibinfo{pages}{517--525}
  (\bibinfo{year}{2000}).
\newblock
  \urlprefix\url{https://www.sciencedirect.com/science/article/pii/S0022459600987218}.

\bibitem{Sasaki1997}
\bibinfo{author}{Sasaki, H.} \emph{et~al.}
\newblock \bibinfo{title}{Structural studies on the phase transition of
  {La$_3$Ni$_2$O$_{6.92}$} at about 550 {K}}.
\newblock \emph{\bibinfo{journal}{J. Phys. Soc. Jpn.}}
  \textbf{\bibinfo{volume}{66}}, \bibinfo{pages}{1693--1697}
  (\bibinfo{year}{1997}).
\newblock \urlprefix\url{https://doi.org/10.1143/JPSJ.66.1693}.

\bibitem{Fernandes2014}
\bibinfo{author}{Fernandes, R.~M.}, \bibinfo{author}{Chubukov, A.~V.} \&
  \bibinfo{author}{Schmalian, J.}
\newblock \bibinfo{title}{What drives nematic order in iron-based
  superconductors?}
\newblock \emph{\bibinfo{journal}{Nature Physics}}
  \textbf{\bibinfo{volume}{10}}, \bibinfo{pages}{97--104,}
  (\bibinfo{year}{2014}).
\newblock \urlprefix\url{http://www.nature.com/articles/nphys2877}.

\bibitem{Proust2019}
\bibinfo{author}{Proust, C.} \& \bibinfo{author}{Taillefer, L.}
\newblock \bibinfo{title}{The {Remarkable} {Underlying} {Ground} {States} of
  {Cuprate} {Superconductors}}.
\newblock \emph{\bibinfo{journal}{Annual Review of Condensed Matter Physics}}
  \textbf{\bibinfo{volume}{10}}, \bibinfo{pages}{409--429}
  (\bibinfo{year}{2019}).
\newblock
  \urlprefix\url{https://doi.org/10.1146/annurev-conmatphys-031218-013210}.



\bibitem{Jin2011}
\bibinfo{author}{Jin, K.}, \bibinfo{author}{Butch, N.~P.},
  \bibinfo{author}{Kirshenbaum, K.}, \bibinfo{author}{Paglione, J.} \&
  \bibinfo{author}{Greene, R.~L.}
\newblock \bibinfo{title}{Link between spin fluctuations and electron pairing
  in copper oxide superconductors}.
\newblock \emph{\bibinfo{journal}{Nature}} \textbf{\bibinfo{volume}{476}},
  \bibinfo{pages}{73--75} (\bibinfo{year}{2011}).
\newblock \urlprefix\url{https://doi.org/10.1038/nature10308}.



\bibitem{Licciardello2019}
\bibinfo{author}{Licciardello, S.} \emph{et~al.}
\newblock \bibinfo{title}{Electrical resistivity across a nematic quantum
  critical point}.
\newblock \emph{\bibinfo{journal}{Nature}} \textbf{\bibinfo{volume}{567}},
  \bibinfo{pages}{213--217} (\bibinfo{year}{2019}).
\newblock \urlprefix\url{https://doi.org/10.1038/s41586-019-0923-y}.

\bibitem{Zhao2022}
\bibinfo{author}{Yuan, J.} \emph{et~al.}
\newblock \bibinfo{title}{Scaling of the strange-metal scattering in
  unconventional superconductors}.
\newblock \emph{\bibinfo{journal}{Nature}} \textbf{\bibinfo{volume}{602}},
  \bibinfo{pages}{431--436} (\bibinfo{year}{2022}).
\newblock \urlprefix\url{https://doi.org/10.1038/s41586-021-04305-5}.




\bibitem{Lee2023}
\bibinfo{author}{Lee, K.} \emph{et~al.}
\newblock \bibinfo{title}{Linear-in-temperature resistivity for optimally
  superconducting {(Nd,Sr)NiO$_2$}}.
\newblock \emph{\bibinfo{journal}{Nature}} \textbf{\bibinfo{volume}{619}},
  \bibinfo{pages}{288--292} (\bibinfo{year}{2023}).
\newblock \urlprefix\url{https://doi.org/10.1038/s41586-023-06129-x}.



\bibitem{JGChen2023}
\bibinfo{author}{Hou, J.} \emph{et~al.}
\newblock \bibinfo{title}{Emergence of high-temperature superconducting phase
  in \LN crystals}.
\newblock \emph{\bibinfo{journal}{Chinese Physics Letters}}
  \textbf{\bibinfo{volume}{40}}, \bibinfo{pages}{117302}
  (\bibinfo{year}{2023}).

\bibitem{polyChengJGARXIV2023}
\bibinfo{author}{Wang, G.} \emph{et~al.}
\newblock \bibinfo{title}{Pressure-induced superconductivity in polycrystalline La3Ni2O7}.
\newblock \emph{\bibinfo{journal}{Phys. Rev. X}} \textbf{\bibinfo{volume}{14}},
  \bibinfo{pages}{011040} (\bibinfo{year}{2024}).
\newblock \urlprefix\url{https://link.aps.org/doi/10.1103/PhysRevX.14.011040}.


\expandafter\ifx\csname url\endcsname\relax
  \def\url#1{\texttt{#1}}\fi
\expandafter\ifx\csname urlprefix\endcsname\relax\def\urlprefix{URL }\fi
\providecommand{\bibinfo}[2]{#2}
\providecommand{\eprint}[2][]{\url{#2}}

\bibitem{LiuZ}
\bibinfo{author}{Liu, Z.J.} \emph{et~al.}
\newblock \bibinfo{title}{Evidence for charge and spin density waves in single crystals of La$_3$Ni$_2$O$_7$ and La$_3$Ni$_2$O$_6$}.
\newblock \emph{\bibinfo{journal}{Sci. China-Phys. Mech. Astron.}} \textbf{\bibinfo{volume}{66}},
  \bibinfo{pages}{217411} (\bibinfo{year}{2023}).
  \newblock \urlprefix\url{https://link.springer.com/article/10.1007/s11433-022-1962-4}.

\bibitem{Eiling}
\bibinfo{author}{Eiling, A.} \&
  \bibinfo{author}{Schilling, J. S.}
\newblock \bibinfo{title}{Pressure and temperature dependence of electrical resistivity of Pb and Sn from 1-300~K and 0-10~GPa-use as continuous resistive pressure monitor ac curate over wide temperature range; superconductivity under pressure in Pb, Sn and In}.
\newblock \emph{\bibinfo{journal}{Journal of Physics F: Metal Physics}} \textbf{\bibinfo{volume}{11}},
  \bibinfo{pages}{623} (\bibinfo{year}{1981}).
  \newblock \urlprefix\url{https://iopscience.iop.org/article/10.1088/0305-4608/11/3/010}.


\bibitem{Mao}
\bibinfo{author}{Mao, H. K.}, \bibinfo{author}{Xu, J.} \&
  \bibinfo{author}{Bell, P. M.}
\newblock \bibinfo{title}{Calibration of the ruby pressure gauge to 800~kbar under quasi-hydrostatic conditions}.
\newblock \emph{\bibinfo{journal}{Journal of Geophysical Research: Solid Earth}} \textbf{\bibinfo{volume}{91}},
  \bibinfo{pages}{4673} (\bibinfo{year}{1986}).
  \newblock \urlprefix\url{https://agupubs.onlinelibrary.wiley.com/doi/10.1029/JB091iB05p04673}.

\end{thebibliography}
\end{document}